\documentclass[manuscript]{acmart}
\usepackage{multicol}
\usepackage{multirow}
\usepackage{soul}
\usepackage{lipsum}
\usepackage{wrapfig}
\usepackage[edges]{forest}
\usepackage{varwidth}

\usepackage{subcaption}
\usepackage{amsthm}
\usepackage{array}
\usepackage{mathrsfs} 
\usepackage{graphicx}
\usepackage{xcolor}
\usepackage{amsmath}
\usepackage{enumitem}
\usepackage{algorithm}
\usepackage{algpseudocode}
\usepackage{float}
\usepackage{array, makecell}
\usepackage{mathtools}  
\usepackage{amsmath}
\usepackage{tabulary}
\usepackage{booktabs}
\usepackage{amsthm}
\usepackage{rotating}
\usepackage{geometry}
\usepackage{longtable}
\usepackage{array}
\usepackage{comment}
\usepackage{tablefootnote}
\usepackage{natbib}
\usepackage{lscape}
\usepackage{afterpage}

\usepackage{pifont}
\usepackage{utfsym}
\newcommand{\xmark}{\ding{55}}
\theoremstyle{plain}
\usepackage{xspace}
\usepackage[utf8]{inputenc}
\usepackage{comment}
\usepackage{svg}
\usepackage{amsfonts}

\newcommand{\refyear}[1]
{\begin{NoHyper} 
\citeyear{#1}
\end{NoHyper}}

\usepackage{hyperref}
\hypersetup{
  colorlinks=true,
  linkcolor=black,
  urlcolor=blue!70!black,
  citecolor=black,
}

\usepackage{xcolor}
\usepackage{url}

\usepackage{glossaries}
\newacronym{cps}{CPS}{Cyber-Physical System}
\newacronym{ids}{IDS}{Intrusion Detection System}
\newacronym{idss}{IDSs}{Intrusion Detection Systems}
\newacronym{iot}{IoT}{Internet of Things}
\newacronym{wsn}{WSN}{Wireless Sensor Networks}
\newacronym{slr}{SLR}{Systematic Literature Review}
\newacronym{mac}{MAC}{Medium Access Control}
\newacronym{pufs}{PUFs}{physical unclonable functions}
\newacronym{rssi}{RSSI}{Received Signal Strength Indicator}
\newacronym{ppf}{PPF}{Probabilistic Packet Flow}
\newacronym{macode}{MAC}{Message Authentication Code}

\newcommand{\cmmnt}[1]{\ignorespaces}

\newcommand{\totalPapers}{2706}
\newcommand{\excludedFromFullAnalysis}{67}
\newcommand{\selectedPapers}{40}
\newcommand{\screeningPapers}{107}

\newcommand\SecTab[1][.15cm]{\hspace*{#1}}
\newcommand{\SecT}{---}

\newtheorem{definition}{Definition}

\newif\ifcomments

\commentsfalse 

\expandafter\newcommand\csname r@tocindent4\endcsname{4in}

\usepackage[usestackEOL]{stackengine}
\edef\tmp{\the\baselineskip}
\setstackgap{L}{\tmp}

\AtBeginDocument{%
  \providecommand\BibTeX{{%
    \normalfont B\kern-0.5em{\scshape i\kern-0.25em b}\kern-0.8em\TeX}}}

\setcopyright{acmcopyright}
\copyrightyear{2024}

\acmBooktitle{Woodstock '18: ACM Symposium on Neural Gaze Detection,
  June 03--05, 2018, Woodstock, NY}
\acmPrice{15.00}
\acmISBN{978-1-4503-XXXX-X/18/06}

\usepackage{color}

\begin{document}

\title{Security Approaches for Data Provenance in the Internet of Things: A Systematic Literature Review}

\author{Omair Faraj}
\authornotemark[1]
\email{ofaraj@uoc.edu}
\affiliation{
  \institution{{\small Internet Interdisciplinary Institute (IN3), Universitat Oberta de Catalunya (UOC), CYBERCAT-Center for Cybersecurity Research of Catalonia}}
  \city{Barcelona}
  \country{{\small Spain}}
  \postcode{08018}
}
\affiliation{
  \institution{{\small SAMOVAR, T\'el\'ecom SudParis, Institut Polytechnique de Paris}}
  \country{France}
}

\author{David Meg\'ias}
\email{dmegias@uoc.edu}
\affiliation{
  \institution{{\small Internet Interdisciplinary Institute (IN3), Universitat Oberta de Catalunya (UOC), CYBERCAT-Center for Cybersecurity Research of Catalonia}}
  \city{Barcelona}
  \country{{\small Spain}}
  \postcode{08018}
}

\author{Joaquin Garcia-Alfaro}
\email{joaquin.garcia\_alfaro@telecom-sudparis.eu}
\affiliation{
  \institution{{\small SAMOVAR, T\'el\'ecom SudParis, Institut Polytechnique de Paris}}
  \streetaddress{19 place Marguerite Perey}
  \city{Palaiseau}
  \country{{\small France}}
  \postcode{91120}
}

\renewcommand{\shortauthors}{Faraj, Meg\'ias, Garcia-Alfaro}
\renewcommand{\shorttitle}{Security Approaches for Data Provenance in IoT}

\begin{abstract}
The Internet of Things (IoT) relies on resource-constrained devices deployed in unprotected environments. \textcolor{black}{Given their constrained nature, IoT systems are vulnerable to security attacks. Data provenance, which tracks the origin and flow of data, provides a potential solution to guarantee data security, including trustworthiness, confidentiality, integrity, and availability in IoT systems.} Different types of risks may be faced during data transmission in single-hop and multi-hop scenarios, \textcolor{black}{particularly due to the interconnectivity of IoT systems, which introduces security and privacy concerns. Attackers can inject malicious data or manipulate data without notice, compromising data integrity and trustworthiness. Data provenance offers a way to record the origin, history, and handling of data to address these vulnerabilities.} A systematic literature review of data provenance in IoT is presented, exploring existing techniques, practical implementations, security requirements, and performance metrics. Respective contributions and shortcomings are compared. A taxonomy related to the development of data provenance in IoT is proposed. Open issues are identified, and future research directions are presented, providing useful insights for the evolution of data provenance research in the context of the IoT.
\end{abstract}



\maketitle

\noindent \textbf{Acknowledgments:} \small{
Authors acknowledge funding from the  PID2021-125962OB-C31 ``SECURING'' project funded by the Ministerio de Ciencia e Innovación, la Agencia Estatal de Investigación and the European Regional Development Fund (ERDF), as well as the ARTEMISA International Chair of Cybersecurity and the DANGER Strategic Project of Cybersecurity, both funded by the Spanish National Institute of Cybersecurity through the European Union -- NextGenerationEU and the Recovery, Transformation and Resilience Plan; and IMT’s CyberCNI chair (Cybersecurity for Critical Networked Infrastructures, cf. \url{https://cybercni.fr/}).}

\section{Introduction}
\label{sec:intro}

\noindent \textcolor{black}{The \gls*{iot} comprises a network of interconnected physical objects, enabling the exchange and collection of data from diverse sources. This infrastructure facilitates data transfer from various environments over an insecure internet connection, where it is processed, managed, and analyzed through various technologies~\cite{AMMAR2018,BOTTA2016}. However, \gls*{iot} increases network vulnerabilities, as it often lacks mechanisms for identity management, data trustworthiness, and access control, making it accessible to attackers. \gls*{iot} applications span across industrial control, healthcare, home automation, car systems, and environmental monitoring~\cite{Lazarescu2017,Manikandan2018,GUBBI2013}. Moreover, data generated by numerous sensor nodes is subjected to in-network processing during transmission to a base station, where provenance, or the tracking of data origin and evolution, becomes critical to ensure data trustworthiness in decision-making processes~\cite{Salmin2015}.} 

\medskip

\noindent \textcolor{black}{To address \gls*{iot} security challenges, data provenance is introduced as a solution to trace and verify the history of data transactions. Initially developed for heterogeneous database systems~\cite{wang1990}, data provenance documents the origin, evolution, and transformations applied to data entities, also known as lineage~\cite{Wang2016}. In \gls*{iot} and other domains, data provenance is defined as the process of recording the history of data origin, evolution, and activities, providing essential security features such as trustworthiness, integrity, and quality assurance~\cite{buneman2001,simmhan2005,ZAFAR2017}. The objective of data provenance is not only to ensure data quality but also to address specific security requirements, including confidentiality, availability, and the prevention of unauthorized access. This paper proposes a taxonomy aimed at providing an in-depth understanding of data provenance in \gls*{iot} by classifying data provenance solutions across three primary aspects: encoding methods, storage mechanisms, and security attack resilience. Provenance encoding refers to techniques used to encode provenance information into \gls*{iot} systems, enabling efficient tracking of data origins. Provenance storage mechanisms focus on various methods to store and retrieve provenance data, while security attacks represent threats against both data and provenance information in \gls*{iot} networks, addressing mechanisms to ensure data integrity and authenticity in adversarial environments. With provenance applied in domains like file systems, databases, cloud computing, and \gls*{iot} networks, this research specifically focuses on its application within the \gls*{iot} and Wireless Sensor Networks (WSN) domains, emphasizing how provenance can be leveraged to enhance data security and authenticity in these environments by providing researchers with a deep understanding of data provenance and offering insights into research challenges, open issues, and future directions.}

\medskip

\begin{figure*}[!b]
  \centering
  \includegraphics[width=0.9\textwidth]{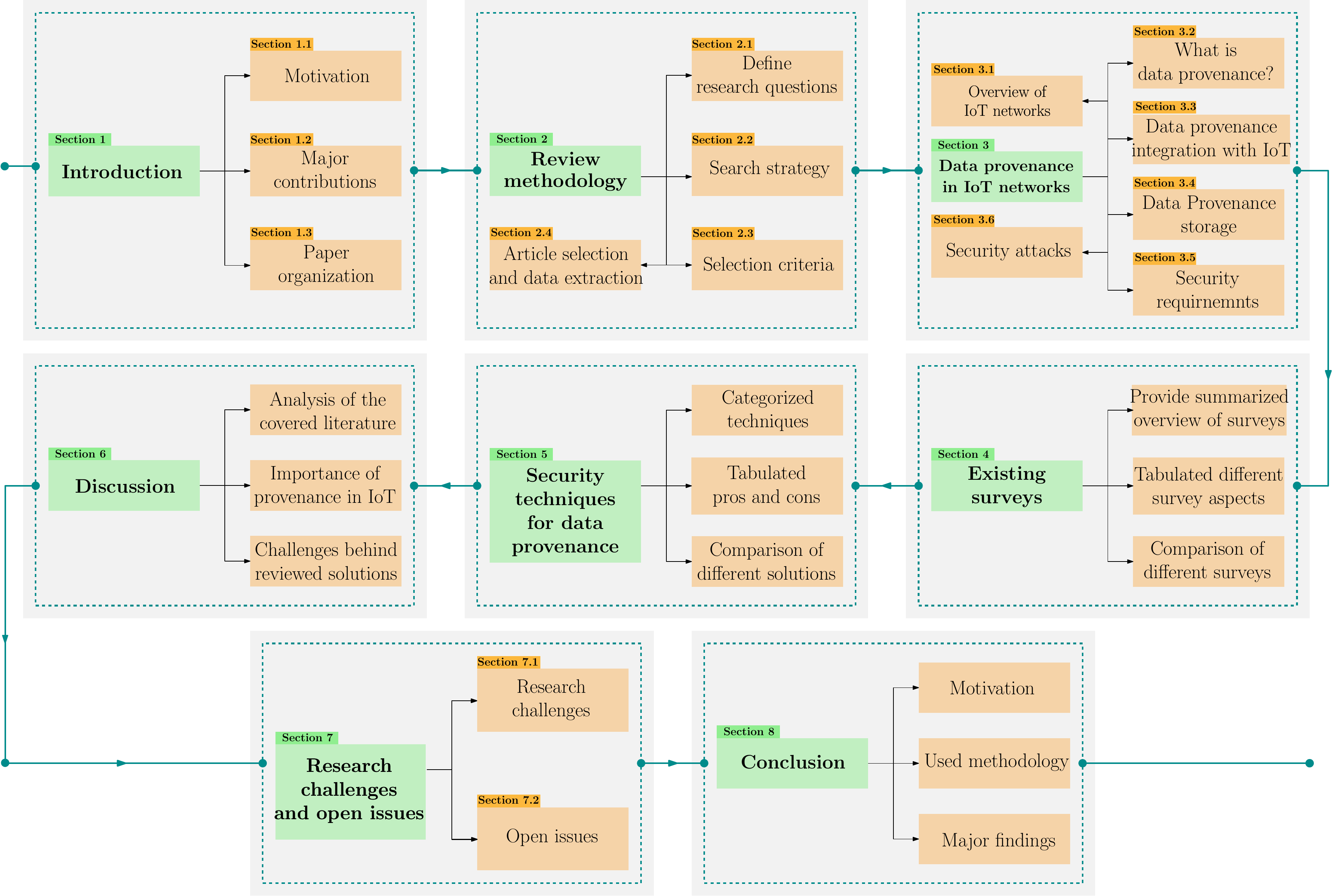}
  \caption{Workflow of this paper.\label{fig:sections}}
\end{figure*}

\noindent \textbf{1.1 \SecTab Motivation \SecT} In \gls*{iot} networks, data provenance provide the capability to ensure data trustworthiness by summarizing the history of ownership and actions performed on collected data from the source node to the final destination. Provenance must be recorded for each collected data packet from source nodes and track the forwarding nodes engaged in the process of data transmission from data origin to destination, but many challenges arise to deploy such a solution. A major challenge is the rapid increase of provenance during the transmission phase in \gls*{iot} networks. Furthermore, there are limitations in data storage capabilities, bandwidth and energy constraints of nodes. Hence, it is critical to address security requirements in such challenging insecure environment in order to securely and efficiently capture provenance information for each packet. Although data provenance has been studied extensively for database management systems it is still not widely addressed in \gls*{iot} networks. Also, the literature still lacks a systematic literature review which introduces the existing proposed data provenance approaches in \gls*{iot} networks and provide a detailed analysis in terms of security requirements, provenance encoding and storing techniques, encountered attacks and performance metrics for the evaluation of proposed methods. We organize this review using the taxonomy shown in Figure~\ref{fig:sections}. Our work aims to provide researchers with the most practical implementations over the past decade to point out open issues and research challenges with in-depth insights for future directions in the field of data provenance in \gls*{iot} based on a \gls*{slr}.  

\medskip

\noindent \textbf{1.2 \SecTab Major Contributions \SecT} The main contributions of our survey are summarized below:
\begin{itemize}
\item We present detailed overview of data provenance and its related concepts, offering essential background knowledge on the security aspects. We also point out potential attacks to data provenance models in \gls*{iot}. We describe the different types of provenance, storage mechanisms and security requirements.
\item We propose an in-depth review of existing techniques for data provenance in \gls*{iot} networks with their advantages, limitations and a set of review attributes through a systematic literature review process.
\item We provide a detailed overview of the encountered attacks and performance metrics used to evaluate security requirements and system efficiency. We propose a taxonomy including different attributes related to the development of data provenance in \gls*{iot}. 
\item We list open issues and research challenges 
on the evolution of data provenance research in \gls*{iot}. 
\end{itemize}

\medskip

\noindent \textbf{1.3 \SecTab Paper Organization \SecT} Section~\ref{sec:methodology} presents the review methodology used to conduct our survey. Section~\ref{sec:background} provides an overview on \gls*{iot} network and data provenance. \textcolor{black}{Section~\ref{storage} provides the mechanisms for data provenance storage systems. The security attacks on IoT networks is introduced in Section~\ref{sec:attacks}.} Section~\ref{sec:surveys} presents existing surveys on data provenance in \gls*{iot}. Section~\ref{sec:techniques} introduces existing data provenance approaches in \gls*{iot} networks, with a comparative analysis. Section~\ref{sec:discussion} provides a discussion. Section~\ref{sec:challenges} presents open issues and research challenges. Section~\ref{sec:conclusion} concludes the paper.

\section{Review methodology}
\label{sec:methodology}
\begin{figure*}[!b]
  \centering
  \includegraphics[width=\textwidth]{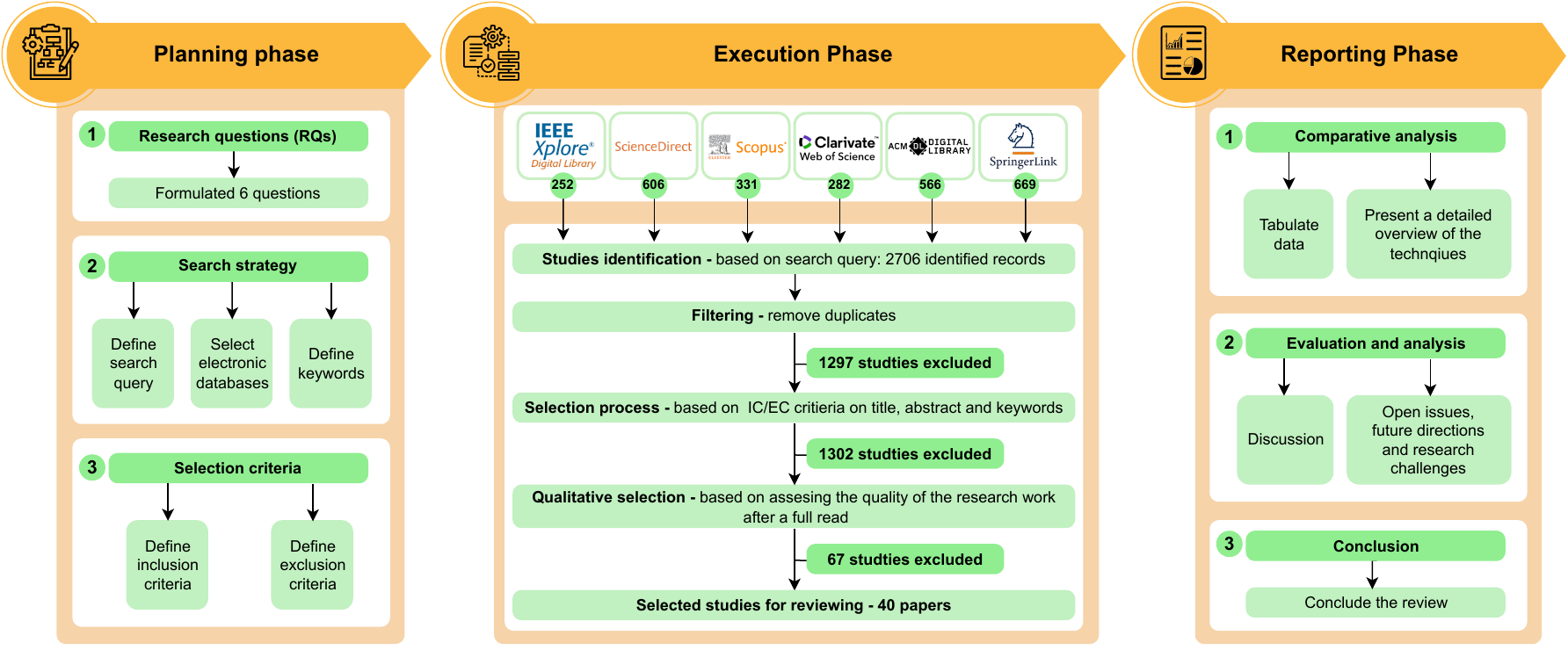}
  \caption{Review methodology phases of our Systematic Literature Review (SLR).\label{fig:methodology}} 
\end{figure*}

A Systematic Literature Review (SLR) is a thorough and well-organized method used to identify, assess, and bring together existing research studies on a specific topic. It follows a precise and transparent process to ensure that all relevant studies are included. By following a systematic and well-defined procedure, this type of study offers many advantages. Firstly, systematic mappings enable a comprehensive overview of the current status of research in a particular area in the state of the art. Secondly, it facilitates the identification of gaps in the literature and provide evidence to guide further research, thereby preventing redundant efforts. Systematic mappings allow for a high-level analysis of all available studies within a specific domain, offering insights into broad research questions concerning the current state of the literature \cite{KITCHENHAM2008,KITCHENHAM2011}. A \gls*{slr} is conducted based on three main phases. \textit{The planning phase}  defines research questions, and provides the search strategy and the criteria to select the relevant state of the art studies. \textit{The execution phase} identifies and selects the required studies according to the previous phase. The \textit{reporting phase} analyzes selected studies and points out open issues and future directions. Our review is based on a \gls*{slr} methodology proposed by~\citet{KITCHENHAM2009}. Figure~\ref{fig:methodology} shows the different phases followed to conduct our review.

\medskip

\noindent \textbf{2.1 \SecTab Research questions \SecT} We use the \gls*{slr} process shown in Figure~\ref{fig:methodology} to provide a comprehensive and exhaustive synthesis of the used methodology. The primary purpose of our review is to investigate the integration of data provenance into \gls*{iot} networks. To plan the review, we formulated six research questions as described in Table~\ref{tab:questions}.

\begin{table*}[!hptb]
\caption{Research questions for the SLR.\label{tab:questions}}
\centering
\resizebox{\textwidth}{!}{
{\footnotesize
\begin{tabular}{c p{5.5cm} p{6.2cm} p{2.2cm}}
\hline
\textbf{\#} & \textbf{Question} & \textbf{Rationale} & \textbf{Where to be addressed?}\\ 
\hline
\textbf{RQ1} & How data provenance is linked to \gls*{iot} networks? & The answer will help understand the main purpose behind using data provenance in \gls*{iot} networks & Section~\ref{sec:background}\\ 
\hline
\textbf{RQ2} & What are the provenance storage techniques, attacks and security requirements when integrating provenance with \gls*{iot}? & The answer will help researchers know the different provenance storing techniques, security attacks against \gls*{iot} system and needed security requirements when implementing a provenance solution & Section~\ref{sec:background}\\ 
\hline
\textbf{RQ3} & How can data provenance security approaches for \gls*{iot} networks be categorized? & The answer will help understanding the classification of proposed methods for the data provenance & Section~\ref{sec:techniques}\\ 
\hline
\textbf{RQ4} & What are the existing practical implementations of data provenance in \gls*{iot}? & The answer will provide researchers with a comprehensive understanding of the shifts in the literature over the past decade. & Section~\ref{sec:techniques}\\ 
\hline
\textbf{RQ5} & What are the advantages and limitations of the proposed techniques for data provenance in \gls*{iot} in the studied literature?  & The answer will help specify the contributions and shortcomings of the studied state of the art & Section~\ref{sec:techniques} and ~\ref{sec:discussion}\\ 
\hline
\textbf{RQ6} &  What are the main research gaps and challenges in the domain of data provenance in \gls*{iot} networks? & The answer will highlight the open issues, and help researchers to identify research challenges and possible future directions. & Section~\ref{sec:challenges}\\ 
\hline
\end{tabular}}
}
\end{table*}

\begin{table*}[!hptb]
\caption{Electronic databases used in the search strategy.\label{tab:databases}}
\centering
{\footnotesize
\begin{tabular}{p{4cm} p{4cm}}
\hline
Database & URL \\ 
\hline
IEEEXplore & \url{http://ieeexplore.ieee.org} \\
Science Direct & \url{http://www.sciencedirect.com} \\
Scopus & \url{http://www.scopus.com} \\
Web of Science & \url{http://www.webofknowledge.com} \\
ACM Digital Library & \url{https://dl.acm.org/} \\
Springer Link & \url{https://link.springer.com/} \\
\hline
\end{tabular}
}
\end{table*}

\noindent \textbf{2.2 \SecTab Search strategy \SecT} To address the research questions, a search process was conducted across six electronic publication databases. The selected databases are shown in Table~\ref{tab:databases}. These databases are widely recognized and widely used in the fields of computer science and engineering, offering extensive coverage of the relevant literature. The automated search process aimed to retrieve relevant studies from these databases~\cite{Chen2010,Malheiros2007}. The set of keywords used to find these studies include the following: \emph{Internet of Things,} \emph{Wireless Sensor Network,} \emph{Data Provenance,} and \emph{Secure Provenance}. The search strategy in the mentioned electronic databases was performed using the following query (and considering synonyms and alternative terms\footnote{The primary keywords were linked together using the logical operator ``AND" to establish the search parameters. Similarly, the variations and synonyms were connected using the logical operator ``OR" to broaden the scope of the search.
}):\\

\begin{center}
{\footnotesize
\begin{tabular}{c}
\hline
\texttt{("Data Provenance" OR "Secure Provenance" OR "Provenance")} \\
\texttt{AND} \\
\texttt{("Internet of Things" OR "IoT" OR "Wireless Sensor Network" OR "WSN")} \\
\hline
\end{tabular}
}
\end{center}

\bigskip

\noindent \textbf{2.3 \SecTab Selection criteria \SecT} To assess the relevance of each primary study in relation to the defined research questions, different selection criteria were applied. The objective was to include studies that have the potential to address the research questions and exclude those that do not contribute to answering them. The criteria is used as a means of determining the suitability of each study for inclusion in the analysis. We have considered several inclusion and exclusion criteria show in Table~\ref{tab:exc-incl}.

\begin{table*}[!hptb]
{\footnotesize
\caption{Inclusion and exclusion criteria used for inclusion in our analysis.
\label{tab:exc-incl}}
\begin{tabular}{p{7cm} p{.25cm} p{6cm}}
\hline
\textbf{Inclusion Criteria} & ~~& \textbf{Exclusion Criteria} \\
\hline
\textit{\textbf{IC1}}: Articles that explores and addresses scenarios, research challenges, and potential opportunities concerning the integration of data provenance and \gls*{iot}. & ~~& \textit{\textbf{EC1}}: The article is not related to data provenance. \newline \textit{\textbf{EC2}}: The article is not focused on data provenance in the domain of \gls*{iot} or \gls*{wsn}. \\
\textit{\textbf{IC2}}: All articles that have original work and used any security data provenance approach and belong to the \gls*{iot} or \gls*{wsn} domain. & ~~& \textit{\textbf{EC3}}: The article serves as an earlier version of a more recent investigation conducted on the same research topic.\\
\textit{\textbf{IC3}}:  Publication years in the range 2013--2023. \newline \textit{\textbf{IC4}}: Articles that are out of the specified time range but have high citations (above 30 citations). \newline \textit{\textbf{IC5}}: Articles written in English. \newline \textit{\textbf{IC6}}: Articles published in journals, conferences and workshops.  & ~~& \textit{\textbf{EC4}}: The article is not written in English. \newline \textit{\textbf{EC5}}: The full text of the article is unavailable. \newline \textit{\textbf{EC6}}: Articles that are repeating in different resources, then, duplicate was excluded.\\
\hline
\end{tabular}
}
\end{table*}

\bigskip

\noindent \textbf{2.4 \SecTab Article selection and data extraction \SecT} Initially, a search was conducted in the chosen electronic databases based on the search strategy, leading to a total of \totalPapers\ papers. The search and selection processes involved considering primary studies published between 2013 and 2023. The reasons for choosing this time range are to make the reference list easier to handle and to avoid older protocols, which usually perform lower than newer approaches, and most approaches are an enhancement of them. Two papers were selected from 2010~\cite{Lim2010} and 2011~\cite{Sultana2011} for being highly cited papers in this field and their direct link to the scope of our research. 

To ensure compatibility with different database engines, slight modifications were made to the search string. The automated search was then conducted on each electronic database, focusing on the title, abstract, and keyword fields. Subsequently, a screening process was carried out on the titles and abstracts to identify and exclude duplicate and irrelevant papers. This screening process led to a reduction in the number of relevant papers to \screeningPapers. The criteria used for exclusion were effective in removing irrelevant studies from this review. Further analysis is conducted on the full texts of these papers, resulting in the exclusion of \excludedFromFullAnalysis\ papers. The remaining ~\selectedPapers\ papers were included in this review. Figure~\ref{fig:methodology} shows the complete identification and selection process of our \gls*{slr}.

To extract data from the chosen primary studies, a data extraction procedure was conducted. This procedure points out items specifically related to the research questions as well as other relevant information. Data extraction was performed on each paper to ascertain the following information: (1) the security technique used to secure provenance, (2) the advantages and limitations of this solution, (3) the approaches and technologies applied to implement provenance encoding, decoding and storing, (4) the metrics used to evaluate the proposed method, and (5) the attacks studied by researchers in the security analysis process of their proposed solution.

\section{Data provenance in IoT networks}
\label{sec:background}
This section provides an overview of \gls*{iot} networks, discussing the attacks on those systems and the necessary security requirements for their design and implementation. Additionally, we introduce the concept of data provenance as a general term and explore its main applications and implementations in \gls*{iot} systems.

\medskip
 
\noindent \textbf{3.1 \SecTab Overview of IoT networks \SecT} The realization of the \gls*{iot} concept in practical terms can be achieved by combining various enabling technologies. In this survey, we direct our attention towards sensor networks, which stand out as a highly used technology across a wide range of applications. Sensor networks have been proposed for numerous scenarios, including environmental monitoring, e-health, transportation systems, military applications, and industrial monitoring , among others. These networks consist of a number of sensing nodes that communicate using a wireless multi-hop model. 
Typically, nodes transmit captured data to a limited number of intermediate forwarding nodes~\cite{ATZORI2010}. In recent years, scientists have conducted extensive research on sensor networks, examining various challenges across different layers of the network protocol stack~\cite{AKYILDIZ2002}. It faces significant vulnerability to attacks due to several factors. Firstly, many of its components are often in stand-by mode for extended periods, making them easy targets for physical attacks. Secondly, the widespread use of wireless communication in the \gls*{iot} makes eavesdropping a simple task. Lastly, most \gls*{iot} components possess limited capabilities in terms of energy and computing resources, which prevents the implementation of complex security schemes. Specifically, trustworthiness and data integrity present major security concerns~\cite{ATZORI2010}. The presence of diverse data sources requires the establishment of trustworthiness for the data, ensuring that only reliable information is taken into account during the decision-making process. In addition, data integrity solutions play an important role in ensuring that any unauthorized modifications to data being transmitted by source nodes or intermediate nodes are detected by the system, preventing adversaries from tampering with the information. For these concerns, new solutions were proposed to ensure the trustworthiness of data, such that any decision process is based on trustworthy data. Data provenance is introduced as one of these effective solution that can be used in constrained networks to ensure trustworthiness of sensed data. Data provenance techniques must consider additional methods to tackle the problem of provenance integrity and data packet integrity. 

\begin{figure*}[!t]
  \centering
  \includegraphics[width=0.9\textwidth]{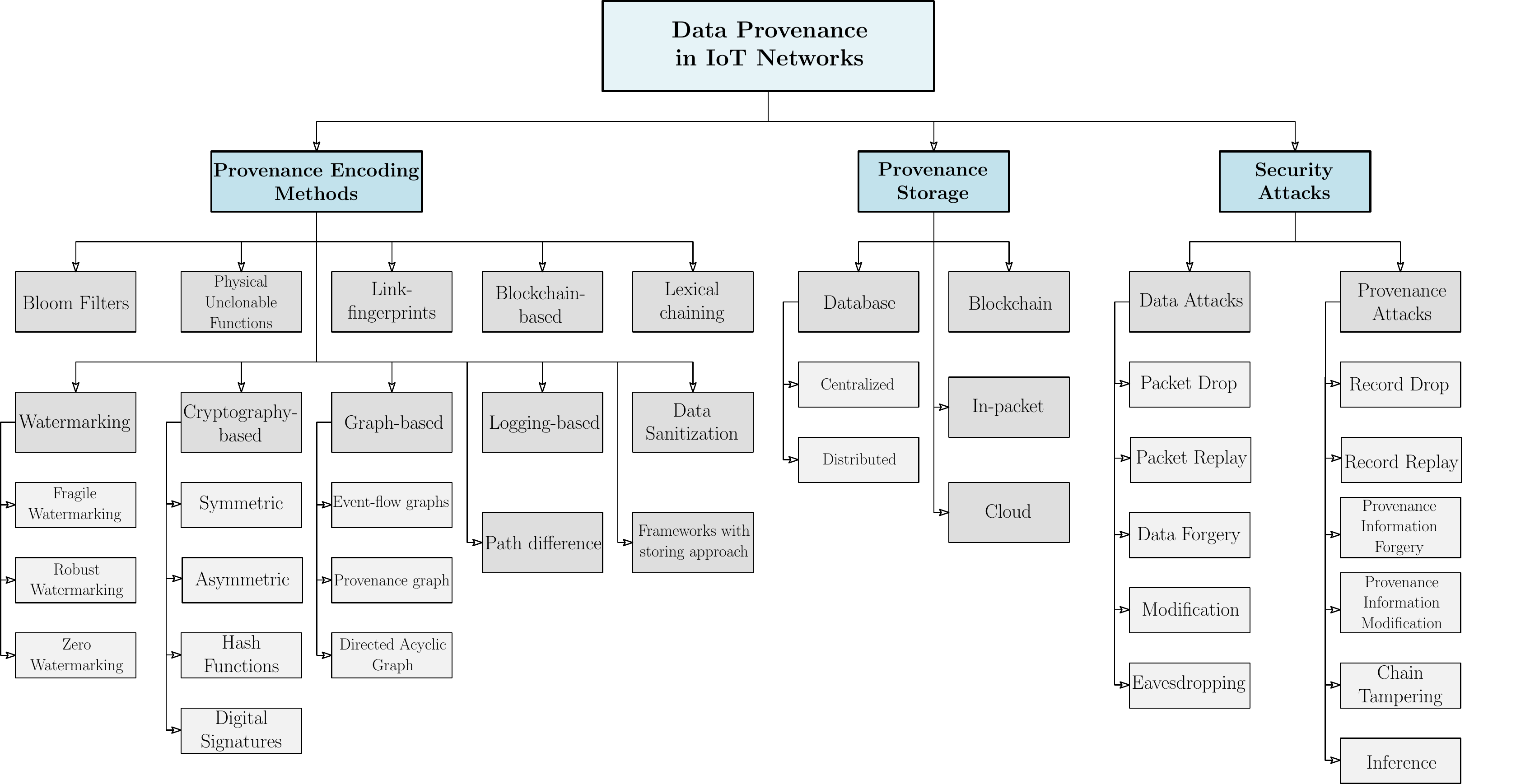}
  \caption{Proposed Taxonomy of Data Provenance in \gls*{iot} networks.}
  \label{fig:taxonomy}
\end{figure*}

\noindent \textbf{3.2 \SecTab What is data provenance? \SecT} The definition of provenance has undergone changes over time, adapting to different application contexts. In database systems, provenance was traditionally referred to as lineage and focused on identifying the source of data resulting from query processing, commonly referred to as data provenance~\cite{pan2023, Moreau2018}. Data lineage was initially formalized by~\citet{cui2000}, specifically in the field of relational databases. It aimed to trace each tuple $"t"$ in the input tables that played a role in generating the output of a query from the database~\cite{james2009}. Provenance, also referred to as pedigree, or genealogy, is a form of metadata that documents the origin and use of a given entity~\cite{simmhan2005,Glavic2007,peter2001}. In the field of information technology, provenance considers data as the counterpart or reflection of an art object. In this field, researchers also define provenance as the complete information about the entire process of data generation and evolution over time. This includes capturing the static origins of data as well as tracking their dynamic changes throughout their transmission process~\cite{Gao2010}. As a summary, provenance holds the capability to provide insights into the what, where, when, how, and why aspects of a given data object. 

\medskip

\noindent \textbf{3.3 \SecTab Data provenance integration with IoT \SecT} The interconnectivity of the \gls*{iot} has brought significant improvements to a wide range of different application scenarios. It has also introduced security and privacy concerns due to data transmission over open networks. Attackers can easily inject malicious data into the transmission path and maliciously manipulate data without any notice from the entities of the networks, thereby compromising the integrity and trustworthiness of the data. \textcolor{black}{For example, in 2016, the Mirai botnet attack exploited unsecured IoT devices, resulting in widespread disruptions of major websites, demonstrating the potential for privacy breaches through compromised devices~\cite{Pankov2023}}. Additionally, the rapid spread of malicious data can cause catastrophic failures. Moreover, the sharing, transmission, and processing of data leads to the risk of user privacy breaches. \textcolor{black}{Unauthorized access to smart home systems can allow intruders to activate webcams and voice-controlled devices or manipulate household appliances like ovens and electric stoves, thereby compromising the security of the entire smart home ecosystem. Such breaches not only jeopardize the functionality of these devices but also pose significant risks to user privacy, as unauthorized individuals may gain access to sensitive personal information or monitor household activities without consent~\cite{David2023}.} Data provenance, which records the origin and processing history of data, presents a potential solution to address these aforementioned issues. Provenance information can record the original source node which produced or forwarded data packet and the operations which the data went through. In many cases, provenance can provide the time and location of the entities that engaged in any operational procedure on data. We provide researchers with a proposed taxonomy for the different aspects of data provenance in \gls*{iot} as shown in Figure~\ref{fig:taxonomy}. 

\begin{figure}[!t]
  \centering
  \subfloat[]{\includegraphics[width=0.3\textwidth]{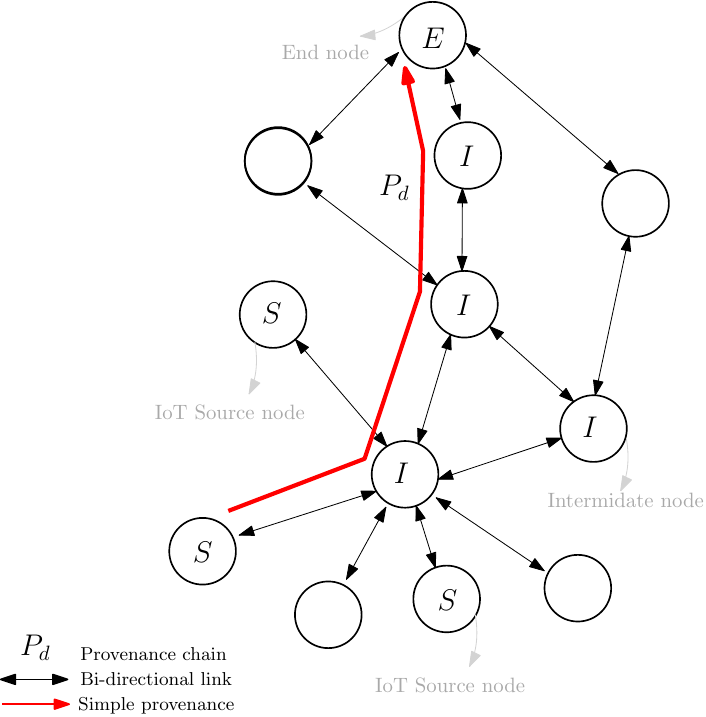}\label{fig:prov_simple}}
  \subfloat[]{\includegraphics[width=0.3\textwidth]{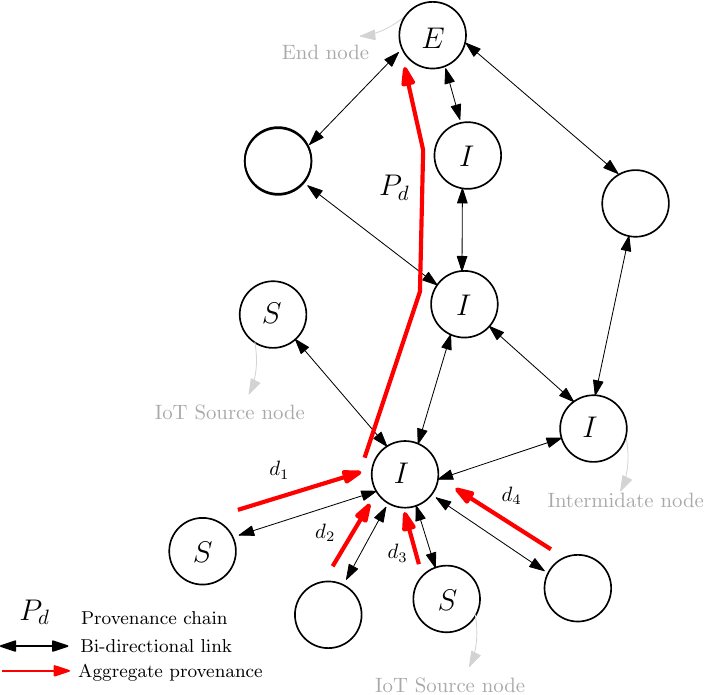}\label{fig:prov_aggregate}}
  \subfloat[]{\includegraphics[width=0.32\textwidth]{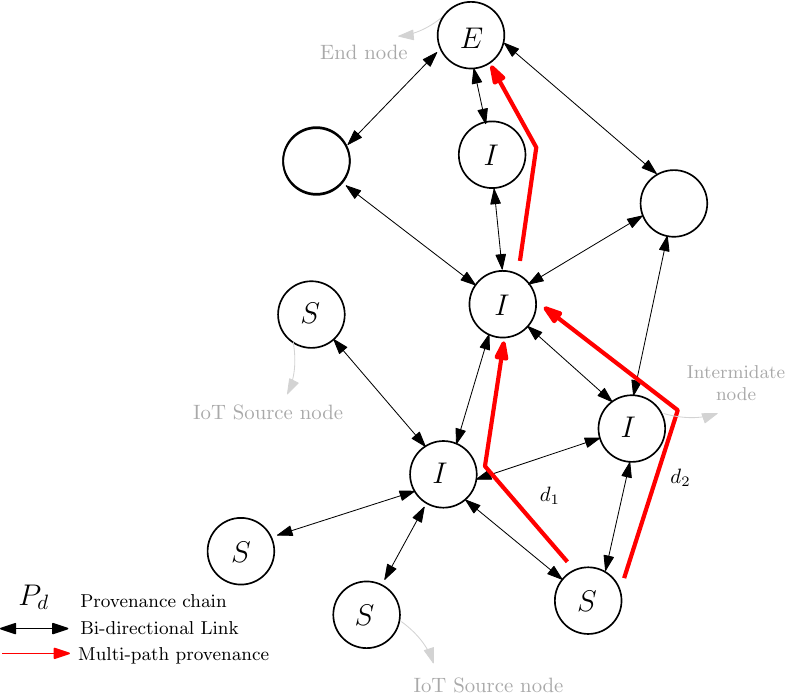}\label{fig:prov_excep}}
  \caption{Data provenance. (a) Simple provenance. (b) Aggregate provenance. (c) Provenance with different data path from one source.}
\end{figure}

The provenance of a data item $d$, denoted as $P_d$, is outlined in Definition~\ref{def:provenance}. $P_d$ records information about the origin and the series of data actions a data item undergoes during its transmission from source to its final destination. There are different types of provenance according to the position of nodes in the network such as simple provenance, as shown in Figure~\ref{fig:prov_simple}; aggregate provenance, as shown in Figure~\ref{fig:prov_aggregate}; and different data path from the same source, as shown in Figure~\ref{fig:prov_excep}. 

\newcommand{\host}{\mathrm{HOST}}

\begin{definition}\label{def:provenance}
    Given a data packet $d$, the provenance $P_d$ is a graph $G(V,E)$  satisfying the following properties: 1) $P_d$ is a subgraph of the sensor network $G(V,L)$; 2) for  $v_i, v_j \in V$, $v_i$ is a child of $v_j$ if and only if $\host(v_i) = n_i$ participated in the distributed calculation of $d$ and/or forwarded the data to $\host(v_j) = n_j$, whereby:
    \begin{itemize}
        \item $N$ = \{$n_i/n_j \,|\, n_i/n_j$ is a network node of identifier is $i,j$\}: a set of network nodes.
        \item $E$ = \{$e_{i,j} \,| \,e_{i,j}$ is an edge connecting nodes $n_i$ and $n_j$\}: a set of edges connecting nodes.
        \item $\host(\cdot)$ assigns a host node to each network node in $V$, where $\host(v_i)$ denotes the host node of the network node $v_i$.
    \end{itemize}
\end{definition}

\noindent \textbf{3.4 \SecTab Data provenance categories --} The main objective when implementing a new data provenance method is to securely store and transmit provenance information. To do so, many challenges in term of security and privacy need to be addressed, which is related to provenance manipulation in collection, storing, and transmission. It is also essential for the designed system to consider the security requirements to overcome such challenging issues~\cite{liu2020,pan2023}. In this context, secure provenance solutions can be divided into the following categories: Cryptography-based, Digital Watermarking, Bloom filters, Physical Unclonable Functions, Fingerprints, Blockchain-based, Frameworks with storing methods, Data Santization, Lexical chaining, Graph-based, Path difference and Logging-based. These categories includes different security techniques that may be combined in some applications to assure data provenance in different \gls*{iot} scenarios. These methods rely on many factors to be developed such as network resources, \gls*{iot} application, needed security requirements, provenance storage approach, energy usage, storage overhead, attack types, and network architecture. The categorization of security techniques for data provenance is shown in the proposed taxonomy in Figure~\ref{fig:taxonomy}. Moreover, a detailed definition and overview on the different categories with their respective techniques is provided in Section~\ref{sec:techniques}. 

\medskip

\noindent \textbf{3.5 \SecTab Security Requirements \SecT} Introducing the challenges of ensuring data trustworthiness in such limited and constrained networks have set a number of security requirements that should be satisfied to have a robust system against different types of attacks. \textcolor{black}{The security requirements were identified through an in-depth analysis of related work (detailed in Table~\ref{tab:requirnemnts} of Appendix~\ref{appendix}) and by considering the specific characteristics and constraints of \gls*{iot} environments, such as limited computational resources and energy efficiency. We also evaluated common attack vectors and vulnerabilities in \gls*{iot} systems to ensure that the requirements address both theoretical and practical security concerns. We can observe that none of the selected papers in the systematic literature review satisfy all of the security requirements necessary for a fully robust provenance system as shown in Table 7. For example, some papers focus on data integrity and confidentiality but lack provisions for ensuring freshness and unforgeability. Others address availability but fail to account for non-repudiation or privacy. This gap highlights the need for a robust solution that encompasses all the key security aspects to effectively protect provenance information in \gls*{iot} networks.} These requirements differ from one scenario to another and are based on the application of the \gls*{iot} system. The requirements are described as follows:

\begin{itemize}
    \item \textit{Data Integrity:}  If the provenance information is altered, it becomes impossible to perform accurate identity management, leading to delayed detection of faulty data propagation. To prevent attackers from manipulating provenance information by selectively adding or removing information, it is important to maintain the integrity of the provenance information~\cite{Hu2020}.
    \item \textit{Confidentiality:} It means that any potential adversary is unable to extract any details about the origin or content of a data packet just by examining the packet's data information and metadata~\cite{Hu2020}.
    \item \textit{Availability:} In general, availability refers to the ability to access data and is often ensured through fault-tolerance mechanisms like data replication across multiple locations. In the context of provenance, availability has been considered a part of integrity. This means that integrity verification is required to confirm, for instance, that provenance data has not been altered~\cite{ragib2009, mohammed2016} or intentionally deleted in a selective manner~\cite{Jamil2018, xinlei2012}.
    \item \textit{Privacy:} \textcolor{black}{It provides extra} guarantees that the sources of this provenance data are not tracked by unauthorized parties. In many cases, the provenance information is more important than the data itself. It requires  protecting privacy of such data. 
    \item \textit{Freshness:} Data freshness refers to the condition where the data is in its most recent and up to date form. This guarantees that no adversary can replay previous information to deceive a gateway or base station. Sensor data measurements are consistently updated, enabling the gateway to rely on the current state of the environmental data, thus preventing reliance on replayed data packets.
    \item \textit{Non-repudiation:} Non-repudiation ensures that a user cannot deny their participation in any activity once the data provenance is recorded. In some cases, both parties need to follow a provenance commitment protocol to ensure mutual non-repudiation. This protocol stops either party from denying their actions or participation in the recorded activities~\cite{asghar2012, backes2016}.
    \item \textit{Unforgeability:} The most critical aspect of a secure provenance system is unforgeability. Unforgeability means that any adversary attempting to alter an existing provenance record or introduce a new forged record will be unable to do so undetected. In other words, unforgeability provides tamper evidence, ensuring the integrity and authenticity of the provenance records~\cite{asghar2012, taha2015, akram2014}.
\end{itemize}

\section{Data provenance storage}
\label{storage}

Efficiently managing the increasing granularity of captured information in provenance, as the number of sensor nodes in a network grows, is a crucial demand. In \gls*{iot}, the provenance expands rapidly due to the increase participation of forwarding nodes. The elements of provenance information vary across applications and methods, depending on the specific requirements fulfilled by the proposed solution. Gathering detailed information about data generation, processing, and forwarding enables the system to ensure diverse security requirements. Thus, it is essential to store this data efficiently to minimize bandwidth, storage, and energy overhead. Provenance information can be stored using four main storage techniques: \textit{local database}, \textit{blockchain}, \textit{in-packet} storage, and \textit{cloud-based}. 

\begin{figure*}[!t]
  \centering
  \subfloat[]{\includegraphics[width=0.46\textwidth]{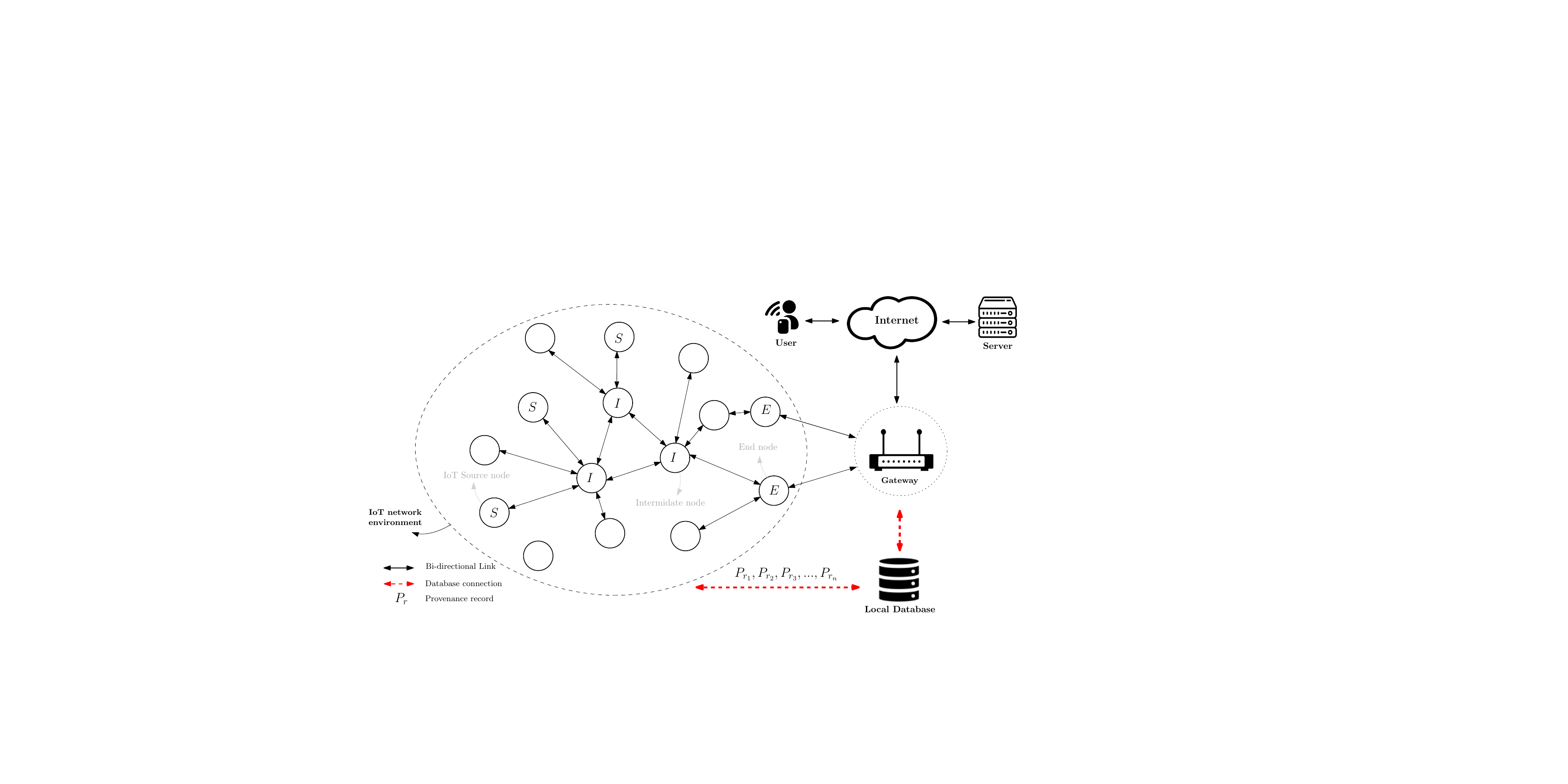}\label{fig:storage_local}}
  \hspace{5mm}
  \subfloat[]{\includegraphics[width=0.27\textwidth]{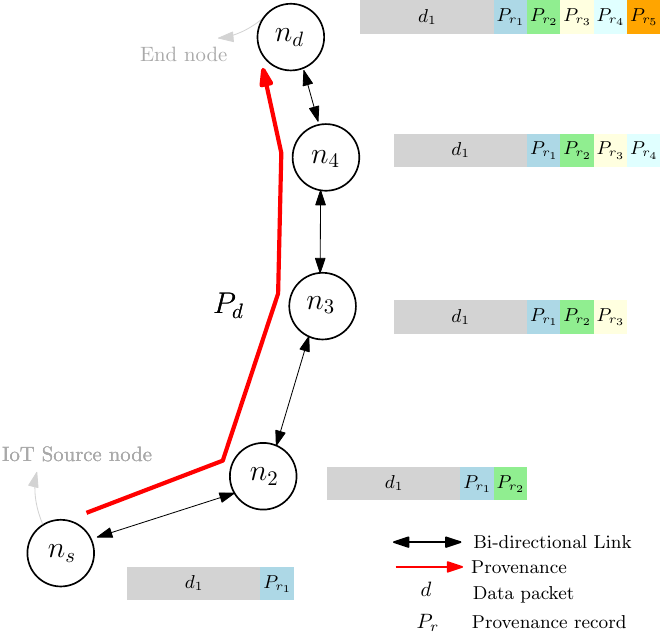}\label{fig:storage_inpacket}}
  \vspace{3mm}
  \hspace{5mm}
  \subfloat[]{\includegraphics[width=0.36\textwidth]{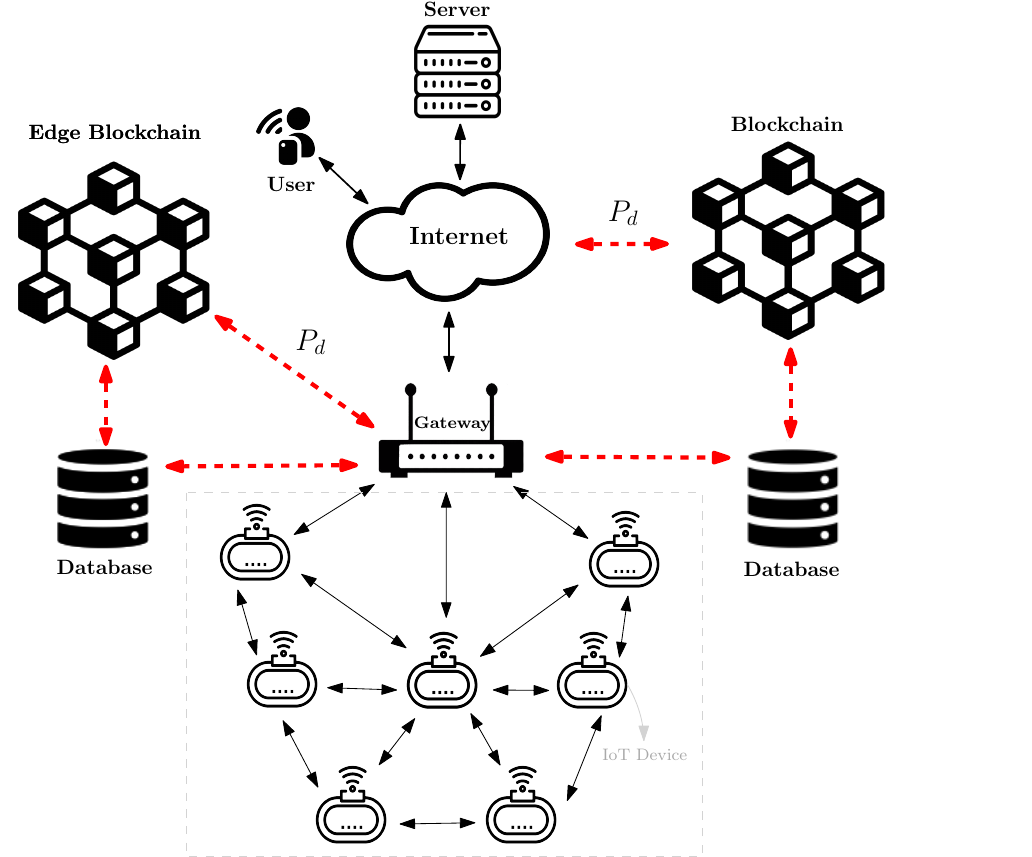}\label{fig:storage_blockchain}}
  \hspace{2mm}
  \subfloat[]{\includegraphics[width=0.46\textwidth]{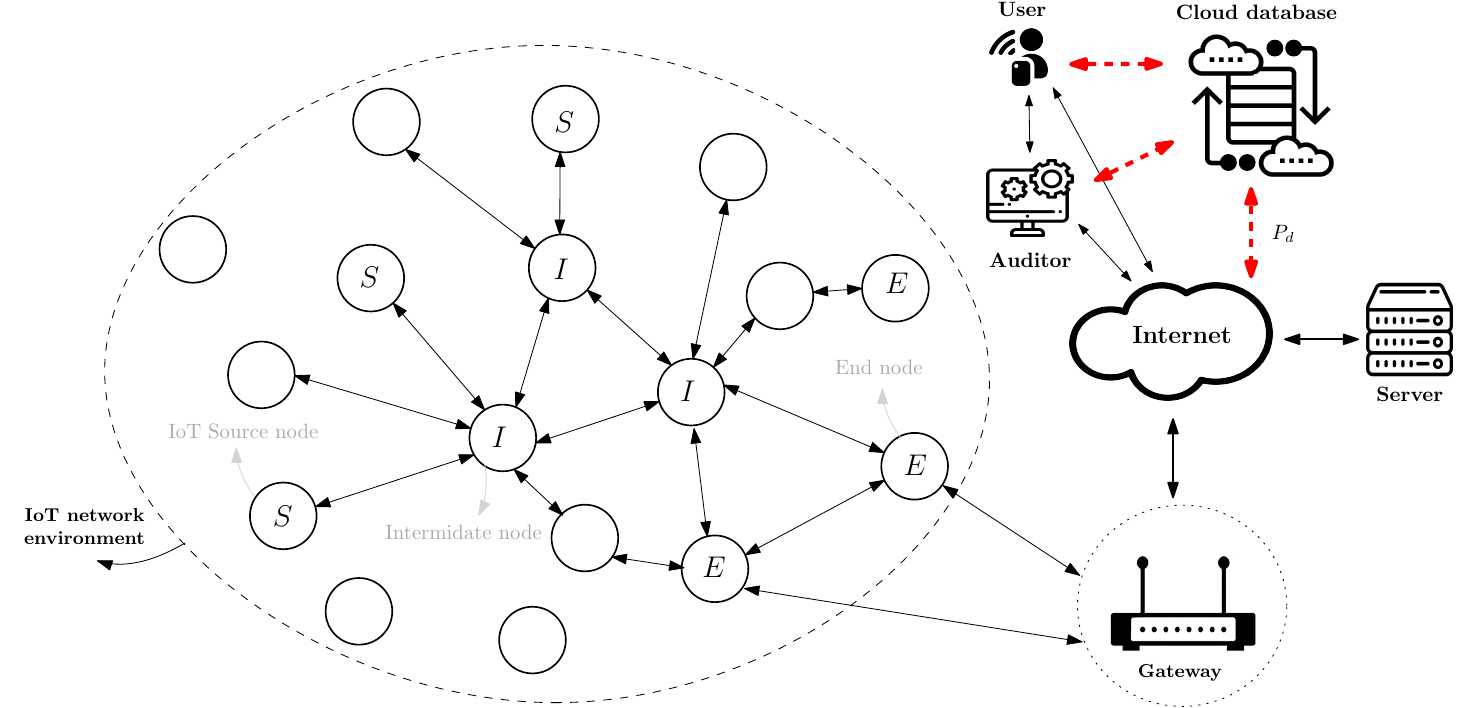}\label{fig:storage_cloud}}
  \caption{Provenance storage mechanisms. (a) Local database. (b) In-packet storage. (c) Blockchain-based. (d) Cloud-based.}
\end{figure*}

\begin{enumerate}
    \item \textbf{Local Database}: In the \textit{local database} system, provenance information is stored either in a distributed or centralized database, depending on the deployed solution's application. \textcolor{black}{However, local storage is typically located at the edge in \gls*{iot} networks for provenance information storage, such as in an edge server, and not directly at resource-constrained \gls*{iot} nodes. This distinction is important, as the storage at the edge does not require an internet connection but is still more capable than individual \gls*{iot} devices. Moreover, as the provenance record grows with the number of forwarding hops, it is not feasible to store it in constrained \gls*{iot} nodes. By storing the information at the edge, it is possible to accommodate larger provenance data without impacting the performance of the \gls*{iot} devices. This approach also enhances scalability, as more resources are available at the edge to handle increasing data volumes~\cite{ZAFAR2017}.} This storage system has challenges regarding storing and querying flexibility, as well as how \gls*{iot} nodes store data information at each forwarding node. Figure~\ref{fig:storage_local} shows an example of the use of a local database in an \gls*{iot} network.
    
    \item \textbf{In-packet}: The second storage technique involves embedding the provenance records within the data packet and maintaining the provenance chain throughout its data path to the final destination as shown in Figure~\ref{fig:storage_inpacket}. As previously mentioned, when data items are processed and transmitted across large-scale systems, the size of the provenance can significantly exceed the size of the data itself. For example, according to the findings of~\citet{jayapandian2006}, in the proposed MiMI system, the provenance associated with data of size 270 MB amounts to approximately 6 GB in size~\cite{jayapandian2006}. This limits the inclusion of numerous information elements in the provenance record, and different provenance systems must selectively retrieve certain provenance information. In many applications, it may not be feasible to ensure security requirements by limiting the provenance information to a very small size which makes it challenging to achieve the required security conditions.
    
    \item \textbf{Blockchain-based}: \textcolor{black}{ \textit{Blockchain} is a }peer-to-peer network's decentralized ledger maintained by all peers, which offers distributed data storage and has been applied to establish data provenance by recording data processes as blockchain transactions~\cite{salman2019}. The large number of data packets generated from different source nodes in the \gls*{iot} network complicates the use of blockchain for storing provenance records, as each record becomes a transaction. This complexity introduces challenges and limitations when applying blockchain for provenance storage. \textcolor{black}{However, in addition to traditional internet-based blockchain solutions, edge-deployed blockchain frameworks, such as Hyperledger Fabric, can also be used to store provenance information closer to the data sources.} An example of connecting a blockchain to an \gls*{iot} network for data provenance is shown in Figure~\ref{fig:storage_blockchain}.

    \item \textbf{Cloud-based}: In the context of \gls*{iot} applications, cloud computing is often used to store provenance records as shown in Figure~\ref{fig:storage_cloud}. Cloud storage provides several advantages for managing and storing provenance records in \gls*{iot}. Cloud storage solutions can easily scale to accommodate the massive volumes of data generated by \gls*{iot} devices. As the number of connected devices increases, cloud platforms can dynamically allocate resources to handle the growing data storage requirements. It is also designed to be compatible with various \gls*{iot} devices and platforms. This ensures that provenance records from diverse sources can be efficiently stored, managed, and retrieved. Integrating cloud storage with existing \gls*{iot} architectures and applications can be complex. Compatibility issues, API variations, and the need for uninterrupted data flow between devices and the cloud may introduce integration challenges. Moreover, \gls*{iot} devices generate huge amount of data, and transferring this data to and from the cloud can introduce latency issues. The efficiency of cloud storage solutions relies on the speed and reliability of data transfer, which can be a challenge, basically in real-time applications~\cite{LI2014,ABIODUN2022,zhang2021,zawoad2018}. An example of connecting a cloud database system to an \gls*{iot} network for data provenance is shown in Figure~\ref{fig:storage_cloud}.
\end{enumerate}

\section{Security Attacks}
\label{sec:attacks}
\gls*{iot} networks are vulnerable to active and passive attacks. The proposed security techniques takes into account these attacks and try to prove their robustness against it through many proposed solutions using different technologies. Provenance information may be more sensitive than the data itself, which makes it a priority for attackers. Attacks on such systems are of two types: \textit{attacks on data} and \textit{attacks on provenance}. Next, we outline some representative attacks that may be perpetrated in an \gls*{iot} environment against data and provenance.

\subsection{Data Attacks} 

\begin{itemize}
    \item \textit{Packet drop attack:} \textcolor{black}{It is a} security threat where an adversary intentionally discards or drops specific data packets within the network. This malicious act disrupts communication between \gls*{iot} devices and can lead to various issues, such as data loss, delayed information delivery, or service unavailability. As \gls*{iot} devices often rely on wireless communication and may operate in resource-constrained environments, packet drop attacks can significantly impact the overall system's functionality and reliability. 
    \item \textit{Packet replay attack:} \textcolor{black}{It is a} type of security attack where an adversary intercepts and records data packets as they travel through the network. The attacker then replays or re-sends these recorded packets at a later time, attempting to deceive the system into thinking the information is fresh and legitimate. This malicious action can lead to various security issues, such as unauthorized access, data manipulation, or false triggering of actions based on the replayed data.
    \item \textit{Data forgery:} \textcolor{black}{It is a} form of security attack where an adversary manipulates or alters the data transmitted between \gls*{iot} devices. The attacker may forge false sensor readings, control commands, or other critical information to deceive the system or cause malicious actions. This type of attack can lead to incorrect decisions, unauthorized access, or compromised system integrity. 
    \item \textit{Data modification attack:} \textcolor{black}{It is a} type of security attack where an unauthorized entity alters or modifies the data being transmitted between \gls*{iot} devices. The attacker may tamper with sensor readings, control commands, or other critical information, leading to incorrect decisions or actions within the system. This can result in operational disruptions, potential risks, or unauthorized access to sensitive data.
    \item \textit{Eavesdrop:} \textcolor{black}{It refers to} the unauthorized interception and monitoring of communication between \gls*{iot} devices. A malicious attacker, also known as an eavesdropper, captures and listens to data packets as they traverse the network. By doing so, the eavesdropper can access sensitive information, such as sensor readings, control commands, or personal data, which can lead to privacy breaches, security vulnerabilities, and potential misuse of the intercepted data.
\end{itemize}

\subsection{Provenance Attacks} 
 
\begin{itemize}
    \item \textit{Provenance record drop attack:} An individual with malicious intent or a group of colluding users can intentionally drop specific provenance records or even the entire set of captured provenance information from a system.
    \item \textit{Provenance replay attack:} An attacker could attempt to deceive the system by replaying a previously recorded provenance record at a later point, thereby manipulating the integrity of the provenance chain.
    \item \textit{Forging provenance attack:} In the context of provenance records, a malicious user or a group of colluding users can engage in forgery by creating false data entries. These forged records can be inserted between legitimate provenance records or added at the end of the existing chain of provenance. The attacks that involve adding false records at the end are commonly known as append attacks. When multiple consecutive adversaries are involved, the process of forging and adding records can be further simplified, as they can only insert forged provenance between themselves.
    \item \textit{Provenance modification attack:} The attacker's objective is to manipulate the provenance record by introducing false data, removing specific information, or modifying existing information to deceive the system.
    \item \textit{Provenance chain tampering:} Provenance records are linked together in a chain to create a complete provenance information of a specific data packet. An attacker's goal is to alter the order of this chain and modify its contents, thereby attempting to manipulate the integrity and reliability of the provenance information.
    \item \textit{Inference attack:} An inference attack compromises the privacy of provenance data, enabling an adversary to deduce sensitive information about the sources and methods used to collect the provenance information.
\end{itemize}

\textcolor{black}{The studied security techniques in the literature, along with the types of attacks each method addresses, are summarized in Table~\ref{tab:attacks} of Appendix~\ref{appendix}. We also provide a taxonomy for these attacks as shown in Figure~\ref{fig:taxonomy}. According to Table~\ref{tab:attacks}, it is evident that the majority of the reviewed papers do not consider most types of attacks. While most papers address forgery and modification attacks, both for data and provenance, they tend to overlook other significant attack vectors such as replay attacks, packet drop, and provenance chain tampering. This limited focus on specific types of attacks suggests that the current security solutions are not fully robust, leaving room for future research to explore more holistic approaches that account for a wider range of potential threats in \gls*{iot} networks.}

\section{Existing surveys}
\label{sec:surveys}

Several review articles address the issue of data provenance in \gls*{iot} networks. Most of these articles focus on the security requirements for data provenance in IoT. In this section, we compare and highlight the importance and need to conduct this review paper. Table~\ref{tab:surveys} provides a comparison between the different review articles based on a number of aspects, including time range (represent the starting and ending years of the conduct review), review methodology (Do the article follow a specified review methodology to conduct the review?), taxonomy (Does the authors provide researchers with a taxonomy that summaries the different aspects used in the review?), security requirements (Does the review use and elaborate the needed security requirements in data provenance approaches?), attacks (Does the review analyze the different approaches based on the security attacks?), performance metrics (Does the review analyze the selected articles based on performance metrics used for the evaluation of any data provenance technique?) and the application domain that the review article covers.  

\citet{Wang2016} provide a deep understanding of provenance schemes, categorizing them into five main categories: distributed schemes, elementary schemes, lossy compression schemes, lossless compression schemes, and block schemes. The survey primarily focuses on \gls*{wsn} networks, examining the techniques used and discussing their respective advantages and disadvantages. One notable limitation of the study is the absence of considerations regarding attacks and performance metrics in relation to the reviewed works. Another survey that categorizes data provenance techniques is presented by~\citet{Hu2020}. The categorization is based on three main technologies logging-based, cryptography-based and blockchain-based technologies. The authors presented the provenance techniques based on the technology used and the requirements to achieve a secure provenance framework. ~\citet{ZAFAR2017} conducted a comparative analysis in their work, presenting a detailed taxonomy of secure provenance schemes. They further perform a discussion of existing schemes, focusing on their strengths and weaknesses. The survey provides a deep understanding of the concept of provenance and its lifecycle, while addressing the necessary security requirements for such systems. The survey does not mention the performance metrics used in the studied methods. ~\citet{alam2021} propose a survey that present and analyze different research methods on three aspects that are provenance management, capture, and analysis. They additionally address the problems of maintaining data integrity, identifying attack chains in trigger-action platforms, and policy compliance in provenance systems of \gls*{iot} environments.

A review on data provenance collection and security in distributed environments is presented by~\citet{Ametepe2021}. The work classifies provenance schemes based on provenance collection schemes, general provenance schemes, and basic provenance schemes. The paper discusses provenance security elements based on integrity, confidentiality and availability, but does not discuss provenance in \gls*{iot} environemnt. ~\citet{gultekin2022} provide an SLR-based methodology on data provenance in \gls*{iot}. The work selects 16 papers for the study. These papers are discussed based on the main subject that the study focus on. The survey lacks comparative analysis based on security requirements and other metrics that are needed to evaluate a provenance system. The most recent survey on data provenance is provided by ~\citet{pan2023}. The study explores the significance of data provenance in the domains of security and privacy. Additionally, the authors outline the foundational principles and models of data provenance, while investigating the mechanisms proposed by previous studies to achieve security objectives. Furthermore, they conduct a review of existing schemes aimed at securing the collection and manipulation of data provenance, commonly referred to as secure provenance and the role of data provenance, specifically in terms of threat provenance. The study covers various application domains beyond \gls*{iot}, including file systems, databases, generic systems, cloud computing, data management, distributed networks, and media data. While the survey examines many of the requirements for provenance systems, it does not specifically address the performance evaluation, deployment technologies and storage methods of provenance techniques in \gls*{iot} environments. Furthermore, it lacks a comparative analysis in this regard. 

The only review paper~\cite{gultekin2022} using a Systematic Literature Review (SLR) methodology for article selection does not address important aspects considered in our work, such as security requirements, security attacks, performance metrics, provenance storage methods, and taxonomy representation. In their review, the authors selected 16 papers, categorized into five main categories: blockchain, security, data flow, data trustworthiness, and privacy. However, the review lacks specific details on data provenance in IoT and does not provide discussions on open issues, research challenges, and future directions. Furthermore, a comparative analysis between the selected articles is not provided. All of the other review articles do not consider all the necessary requirements and aspects essential for analyzing security techniques proposed for data provenance in \gls*{iot}. Our survey focuses on \gls*{iot} environments and provides an extensive comparative SLR review taking into account all the missing aspects from the discussed literature.

\begin{table*}[t]
\begin{scriptsize}
\renewcommand{\arraystretch}{1.7}
\caption{Comparison with other review papers in the literature.\label{tab:surveys}}
\begin{center}
\begin{tabular}{>{\centering}m{2.5cm} | c | c c c c c c p{2.2cm}}
\hline
\multirow{2}{*}{\textbf{Reference}} & \multirow{2}{*}{\textbf{Year}} & \textbf{Time} & \textbf{Review}  & \multirow{2}{*}{\textbf{Taxonomy}} & \textbf{Security} & \multirow{2}{*}{\textbf{Attacks}} & \textbf{Performance}  &   \textbf{Application} \\ 

 &  & \textbf{Range} & \textbf{Methodology}  &  & \textbf{Requirements} &  & \textbf{Metrics}  &   \textbf{Domain}\\ 
\hline
\hline
~\citet{Wang2016} & 2016 & 2006-2016 & \xmark  & \xmark & \usym{1F5F8} & \xmark & \xmark  &   \gls*{wsn}\\
\hline

~\citet{ZAFAR2017} & 2017 & 2007-2017 & \xmark  & \usym{1F5F8} & \usym{1F5F8} & \usym{1F5F8} & \xmark & Cloud computing, \gls*{wsn}, \gls*{iot}, smartphones\\

\hline
~\citet{Hu2020} & 2020 & 2010-2018 & \xmark  & \xmark & \usym{1F5F8} & \xmark & \xmark  &  \gls*{iot}\\
\hline
~\citet{alam2021} & 2021 & 2005-2020 & \xmark  & \xmark & \usym{1F5F8} & \xmark & \xmark  &  \gls*{iot}\\
\hline
~\citet{Ametepe2021} & 2021 & 2010-2018 & \xmark  & \usym{1F5F8} & \usym{1F5F8} & \xmark & \xmark  &  Distributed Environments\\
\hline
~\citet{gultekin2022} & 2022 & 2012-2022 & \usym{1F5F8}  & \xmark & \xmark & \xmark & \xmark  &  \gls*{iot}\\
\hline

~\citet{pan2023} & 2023 & 2009-2022 & \xmark  & \xmark & \usym{1F5F8} & \usym{1F5F8} & \xmark  &  General\\

\hline

This review & 2024 & 2013-2023 & \usym{1F5F8}  & \usym{1F5F8} & \usym{1F5F8} & \usym{1F5F8} & \usym{1F5F8}  &  \gls*{wsn}/\gls*{iot}\\
\hline
\end{tabular}
\end{center}
\end{scriptsize}
\end{table*}

\section{Security techniques for data provenance}
\label{sec:techniques}
Data provenance is a concept that is applied in many fields of study. It is uniquely defined by each application domain~\cite{Park2008}. Data provenance in \gls*{iot} networks serves to guarantee data trustworthiness by collecting the lineage of ownership and actions executed on collected data from the source node to its final destination. It is essential to record provenance for every data packet generated from source nodes and trace the involvement of forwarding nodes throughout the data transmission process, but deploying such a solution presents numerous challenges. One significant challenge is the rapid increase in provenance data during the transmission phase in \gls*{iot} networks. Additionally, limitations arise from the constraints imposed by data storage capabilities, bandwidth, and energy consumption of nodes~\cite{Lim2010}. Data provenance methods establish user trust in the received information by validating its origin, ensuring that data packets were generated by the designated and authorized \gls*{iot} node at the specified time and location~\cite{Bertino2018}. Provenance can be described as a chain of nodes which traverses from source to destination as shown in Figure~\ref{fig:network}. 

The idea of data provenance has been applied by many researchers for identifying the source of data, trace ownership to ensure data authenticity, and evaluate trustworthiness. \textcolor{black}{We classify and study the provided solutions based on the following main technologies: Watermarking, Data Sanitization, Lexical Chaining, Path Difference, Logging-based Techniques, Bloom Filters, Fingerprints, Frameworks using Storing Methods, Graph-based Provenance, Cryptography-based Techniques, Physical Unclonable Functions, and Blockchain-based Solutions. In this section, the provenance encoding techniques are ordered by increasing overhead, beginning with lighter methods and progressing to those with higher resource demands.} In Appendix~\ref{appendix}, we provide a detailed comparison of the selected papers based on a number of \textit{protocol metrics}, \textit{performance metrics}, \textit{security requirements}, and \textit{attacks}.

\noindent \textbf{5.1 \SecTab Watermarking \SecT} Watermarking is a widely known advancement in the field of \gls*{wsn} security. It serves the purpose of identifying any alterations made to sensory data, effectively preventing unauthorized interception~\cite{megias2010,araghi2023}. The two main categories of digital watermarking are fragile watermarking and robust watermarking, each offering different anti-attack properties. Fragile watermarks become undetectable when data is modified, whereas robust watermarks can withstand various forms of distortion~\cite{cox2007,megias2021}. Watermarking is being used in many security applications and is introduced in some proposed data provenance works. \textcolor{black}{As shown in Table~\ref{tab:comparison_sanitization}, one work discusses watermarking as a solution for data provenance protection.} ~\citet{Sultana2011} develop a technique for data provenance aimed at identifying malicious packet dropping attacks. The technique depends on the timing characteristics between packets after the process of embedding provenance information. Based on the distribution of inter-packet delays it detects the packet loss. Then, it identifies the presence of an attack and localizes the malicious node or link.

\noindent \textbf{5.2 \SecTab Data Sanitization \SecT} There are situations where privacy concerns prevent some information from being shared inside the provenance chain. This is where the method of data sanitization is presented. Data sanitization assists in hiding some important information which could compromise privacy. Data sanitization is mainly used to make sure that data is consistent, accurate, and dependable, such that it can be used for reporting, analysis, and data-hiding for privacy preservation. ~\citet{Lomotey2019} present two approaches for device and data verification in an \gls*{iot} network to achieve trust and privacy preservation. The first approach enables devices to subscribe and reveal their metadata to allow for data packet tracing without compromising privacy. To achieve this, a data sanitization method for hiding sensitive information is applied. Users and devices have the option to label attributes within the provenance data as non-shareable with other devices in the chain. In the second approach, the authors modeled the entire peer-to-peer \gls*{iot} network as a graph network. Data origins are verified using Floyd's algorithm between different interconnected nodes. This work ensures transparency, traceability, and privacy preservation through both proposed approaches. \textcolor{black}{The technique is presented in Table~\ref{tab:comparison_sanitization}.}

\noindent \textbf{5.3 \SecTab Lexical Chaining \SecT} Semantically similar words or phrases are linked together by lexical cohesiveness. The related elements can be linked together to create Lexical Chains after all cohesive relationships have been determined forming a conceptually accurate building blocks in a variety of natural language processing systems~\cite{biemann2013}. This is how a typical chain may be written: 
$$
\text{Device} \rightarrow \text{Type} \rightarrow \text{Owner} \rightarrow Communicated
$$
would be feasible to track and categorize the communication channels between different \gls*{iot} devices as well as their previous interactions using this method, which may be used for machine-to-machine communication. In their work, ~\citet{Lomotey2018} first emphasize the use of provenance through the design of algorithms to confirm the origin of IoT-based data, and then they propose developing completeness techniques through visual analytics to trace data packets through the complete route in an \gls*{iot} network. They present an improved system provenance mechanism in their study to achieve traceability. The technique is based on an associative logic to decide all connections established in a machine-to-machine communication between \gls*{iot} nodes. To create links between device communication and data propagation to the $n$-th degree, a statistical lexical chaining based on the Adjusted Rand Index (ARI) is suggested as an alternative to knowledge-based methodology. Because data propagation pathways and object-to-object communications can be identified, the proposed \gls*{iot} architecture makes traceability easier. Based on their findings, the suggested system demonstrates that, in terms of identifying linkability, unlinkability, and availability, the ARI is more accurate than the knowledge-based methodology. Additionally, visual analytics is offered to give a clearer understanding of interconnection in \gls*{iot} nodes using visualization graphs such as HyperTree Graph, the Weighted Graph, and the combination of SpaceTree and RGraphs. \textcolor{black}{These methods are presented in Table~\ref{tab:comparison_sanitization}.}

 \noindent \textbf{5.4 \SecTab Path Difference \SecT} Path difference is the sum of an indicator variable \textcolor{black}{which represents whether} the actual packet path for the next hop is the same as the parent node of this packet along the routing path. Parent information can be used in reference packets for path reconstruction if the path difference value is equal to zero. If not, more time will be required to record the real forwarders. In order to minimize the message overhead, it is better to have a minimal path difference value. In big-scale sensor networks having lossy links and complex routing dynamics,~\citet{gao2013} mention the ineffectiveness of current path reconstruction techniques. The authors present Pathfinder, a cutting-edge path reconstruction method. Pathfinder takes advantage of temporal correlation between a group of packet paths at the node side and effectively compresses the route information with path variation; at the PC side, Pathfinder determines packet routes from the compressed data and uses smart path speculation to reassemble the data packet paths with a high reconstruction ratio. With the use of extensive simulations and traces from a large real-world sensor network, they construct Pathfinder and evaluate its performance against the two most similar approaches. Results indicate that Pathfinder performs noticeably better than MNT~\cite{keller2012} and PathZip~\cite{lu2012} in a number of network setups. Their findings show that Pathfinder achieve both low transmission overhead and high reconstruction ratio. \textcolor{black}{An overview of the proposed technique is shown in Table~\ref{tab:comparison_sanitization}.}

\noindent \textbf{5.5 \SecTab Logging-based Techniques \SecT} Concerns over creating appropriate forensic investigation models were highlighted by cybercrime occurrences seen in \gls*{iot} networks. Every attack on an \gls*{iot} network leaves behind some evidence of it, but the primary difficulty is locating, gathering, and correlating that evidence for accurate forensic analysis. In order to provide answers to specific investigation queries, it can sometimes be rather challenging to determine the linkages between discrete information gathered from \gls*{iot} devices and network level activity. To tackle this issue, forensic investigators can benefit from provenance logging. Over time, network provenance generates a vast amount of information. \gls*{iot} devices reduce this cost by using template systems, which require them to log just basic network characteristics as the traditional device logging. 

~\citet{SADINENI2023} propose a template-based provenance approach, ProvLink-IoT, to provide reliable forensic analysis in \gls*{iot} networks\textcolor{black}{, as presented in Table~\ref{tab:comparison_sanitization}.} ProvLink-IoT is developed to analyze link-layer attacks. Many open source tools are used to implement the system in a simulated environment to approve its robustness in correlating evidence that is found in link-layer provenance. Based on provenance logs and network data gathered from the network, traceability graphs are generated in both normal and attack scenarios. The approach is studied using 6TiSCH network stack. To analyze the effects and conduct forensic analysis, the authors applied three link-layer attacks to the Time Slotted Channel Hopping (TSCH) and 6top (i.e. operational sublayer) layers of the 6TiSCH network. For performance evaluation, they used storage overhead and provenance growth rate.

\begin{table}[!htbp]
\begin{scriptsize}
\renewcommand{\arraystretch}{1.5}
\begin{center}
\caption{Overview of selected studies using Watermarking, Sanitization, Lexical Chaining, Path Difference and Logging-based methods for provenance encoding in IoT networks.\label{tab:comparison_sanitization}}
\centering
\resizebox{\textwidth}{!}{\begin{tabular}{c | c >{\centering}m{2.5cm} c >{\centering}m{2.5cm} c m{4cm} m{4.3cm}}
\hline
\multirow{2}{*}{\textbf{Reference}} & \multirow{2}{*}{\textbf{Year}}   & \textbf{Provenance Encoding} & \textbf{Provenance Storage} & \multirow{2}{*}{\textbf{Application}} & \textbf{Security} & \multirow{2}{*}{\textbf{Pros}} & \multirow{2}{*}{\textbf{Cons}} \\ 

&  & \textbf{Method} & \textbf{Method} &  & \textbf{Analysis} &  &  \\ 
\hline
\hline
\multicolumn{8}{|c|}{\textbf{Watermarking}} \\
\hline
~\citet{Sultana2011} & \refyear{Sultana2011} & Watermarking based & inter packet delays & sensor network & \xmark & High detection accuracy, energy efficiency & Drastic increase in provenance size as number of nodes increases  \\
\hline
\multicolumn{8}{|c|}{\textbf{Data Sanitization}} \\
\hline
~\citet{Lomotey2019} & \refyear{Lomotey2019} & Data sanitization & Local database & Sensor network & \xmark & Hides sensitive data that users do not want to share with others by tagging attributes in the provenance information. & Provenance information is not secure. Storing and querying provenance information is not defined. Integrity is not ensured. \\
\hline
\multicolumn{8}{|c|}{\textbf{Lexial Chaining}} \\
\hline
~\citet{Lomotey2018} & \refyear{Lomotey2018} & Associative rules and statistical lexical chaining & Centralized database (CouchDB) & Devices with Bluetooth in a machine-to-machine (M2M) scenario & \xmark & During machine-to-machine communication, it is possible to track and categorize the communication routes taken by various \gls*{iot} devices as well as their previous connections. & Lexical chains are not securely communicated between the different entities and the database. \\
\hline
\multicolumn{8}{|c|}{\textbf{Path Difference}} \\
\hline
~\citet{gao2013} & \refyear{gao2013} & Path difference and speculation & In-packet/ PC database & \gls*{wsn} & \xmark & Lightweight approach that does not require complex computation at sensor nodes. In case of inability to record path difference, a path speculation method can reconstruct the routing path. & The path field container is limited to a number of bits that cannot be exceeded. When the path difference is large and exceeds this limit, the path cannot be recorded completely. \\
\hline
\multicolumn{8}{|c|}{\textbf{Logging-based}} \\
\hline
~\citet{SADINENI2023} & \refyear{SADINENI2023} & Provenance logs and network traffic & Centralized database & Link-Layer Forensics in \gls*{iot} & \xmark & Detects network and link layer attacks in \gls*{iot} networks. Detects stealthy attacks. & Provenance information is not securely transmitted through transmission channel. The stored provenance information can be altered. Multi-hop provenance path construction is not studied. Due to rapid increase in provenance information, there is a storage overhead at the database. \\
\hline
\end{tabular}}
\end{center}
\end{scriptsize}
\end{table}

\noindent \textbf{5.6 \SecTab Bloom Filters \SecT} A compact data structure, using hashing, enables rapid verification of item presence within the structure. Provenance information is embedded within the generated structure and the bloom filter is transmitted along with the data. By employing this method, the original information remains inaccessible to potential adversaries. Due to the variability of the bloom filter value from one packet to another, establishing a connection or association between previous bloom filters and recent data becomes challenging. \textcolor{black}{An overview of the selected papers in this category are shown in Table~\ref{tab:comparison_bloomfilters}.} In their work, ~\citet{Salmin2015} present a secure provenance scheme for \gls*{wsn}. Their approach involves embedding provenance information into a Bloom Filter, which is transmitted alongside the data. This scheme effectively addresses the challenges posed by resource constraints in \gls*{wsn}. It requires a single channel for data and provenance transmission. The scheme overlooks data integrity and only focuses on studying the packet drop attack. Additionally, ~\citet{SIDDIQUI2019} present a data provenance technique for \gls*{iot} devices that employs Bloom Filter and attribute-based encryption. This approach presents challenges as \gls*{iot} devices typically have limited memory capacity, making it impractical to store extensive provenance information. Furthermore, the technique is vulnerable to physical attacks, whereby an attacker can easily manipulate the stored provenance information within an \gls*{iot} device.

~\citet{harshan2020} introduce a method for embedding provenance information known as the Deterministic Double-Edge (DDE) embedding mechanism, which is based on Bloom filters~\cite{lu2012,shebaro2012}. In this approach, a relay node embeds information about both edges connected to it: the edge through which it receives a packet and the one through which it sends the packet. This means that a node can effectively cover two edges in the path with a single action, reducing the need for the next node in the path to update the provenance. This results in decreased packet delay. Authors show that the hop-counter within the provenance segment can also be used to coordinate the skipping strategy among nodes. Additionally, they propose upper bounds on the error rates of the DDE embedding technique, which are expressed as functions of the network's node count, the number of hops, the size of the Bloom filter, and the number of hash functions computed by each node.

\begin{table}[!htbp]
\begin{scriptsize}
\renewcommand{\arraystretch}{1.5}
\begin{center}
\caption{Overview of selected studies using bloom filters-based methods for provenance encoding in IoT networks.\label{tab:comparison_bloomfilters}}
\centering
\resizebox{\textwidth}{!}{\begin{tabular}{c | c >{\centering}m{2.5cm} c >{\centering}m{2.5cm} c m{4cm} m{4.3cm}}
\hline
\multirow{2}{*}{\textbf{Reference}} & \multirow{2}{*}{\textbf{Year}}   & \textbf{Provenance Encoding} & \textbf{Provenance Storage} & \multirow{2}{*}{\textbf{Application}} & \textbf{Security} & \multirow{2}{*}{\textbf{Pros}} & \multirow{2}{*}{\textbf{Cons}} \\ 

&  & \textbf{Method} & \textbf{Method} &  & \textbf{Analysis} &  &  \\ 
\hline
\hline
\multicolumn{8}{|c|}{\textbf{Bloom Filters}} \\
\hline
~\citet{Salmin2015} & \refyear{Salmin2015} & in-packet Bloom filters & in-packet chain & \gls*{wsn} & \usym{1F5F8} & Simple encoding scheme. Reduces the size of provenance length. Scalable for a high number of nodes. & Provenance information is not sufficient (only node ID). Sends the complete data path within the packet.  \\
~\citet{SIDDIQUI2019} & \refyear{SIDDIQUI2019} & Bloom Filters with cryptographic mechanisms & in-packet & N/A & \xmark & The computation of the partial digital signature is carried out by the \gls*{iot} node, while the more resource-intensive calculations are offloaded and performed by the edge node. Enhanced storage capabilities.  & Fast increment of provenance size. Complete provenance is appended to the data packet. \\
~\citet{harshan2020} & \refyear{harshan2020} & Bloom Filter & In-packet & Raspberry Pi network & \xmark & At most half the nodes in the path modify the provenance. Reduction in the delay on packets. & Storage overhead in large-scale networks as the number of hops increases. \\

\hline
\end{tabular}}
\end{center}
\end{scriptsize}
\end{table}

\medskip
 
\noindent \textbf{5.7 \SecTab Fingerprints \SecT} Fingerprints refers to unique identifiers or signatures assigned to the communication links between \gls*{iot} devices. These fingerprints are typically generated based on various characteristics or attributes of the link, such as received signal strength, latency, packet loss, or other network metrics. The homomorphic feature of public-key cryptography is exploited by numerous anonymous fingerprinting systems. These approaches enable the user's fingerprint to be embedded in an encrypted domain (using public key) so that only the user may access the decrypted fingerprinted material by using the private key~\cite{megias2015}. Data provenance solutions have been developed by analyzing and comparing link fingerprints generated from different attributes. \textcolor{black}{The selected studies based on this technique is shown in Table~\ref{tab:comparison_fingerprints}.} In their work, ~\citet{syed2014} present a method for enhancing the security of data provenance in bodyworn medical sensor devices. They achieve this by using the spatio-temporal characteristics of the wireless channels used for communication by these devices. This solution allows two parties to create highly similar link fingerprints, which uniquely link a data session to a specific wireless connection. This link enables a third party to later verify transaction details, especially the specific wireless link through which the data was transmitted. To validate this approach, experimental testing was conducted using MicaZ motes running TinyOS, and the results show an improvement in energy efficiency. Additionally, the proposed technique generates provenance information on each session, reducing the use of cryptographic methods. Using different fingerprint method, Alam and Fahmy~\cite{ALAM2014} propose a provenance encoding and construction method that adapts three encoding schemes: juxtaposition of ranks, prime scheme, and Rabin fingerprints. The method is referred to as \gls*{ppf}. In the first scheme, provenance is constructed based on the rank of the node instead of the node ID, which requires fewer bits for encoding. Assuming that the packet contains a specific space to hold the identities of $m$ nodes, a counter of $\log(m)$ bits is used to track the embedded ranks in the packet. The second scheme is based on prime multiplication, which embeds more node IDs using the same number of bits compared to the previous one. This method uses prime numbers as node IDs, and their multiplication results in encoding a set of IDs that can be uniquely factorized. In their third method, a data path traversed by a packet is a sequence of bits that represents the IDs of nodes on that path. The fingerprint of this sequence of bits is transmitted instead of the actual sequence. ~\citet{Kamal2018} introduce a lightweight protocol for a multi-hop \gls*{iot} network, aiming to ensure both data security and the establishment of data provenance. The protocol uses link fingerprints derived from the \gls*{rssi} of \gls*{iot} nodes within the network. The protocol achieves data provenance by appending the encoded link fingerprint to the data packet header as it traverses each node. After receiving the packet, the server decode the packet header in a sequential process. However, provenance data expands rapidly, which requires transmitting a large amount of provenance information to data packets, thereby increasing the bandwidth overheads.

\begin{table}[!htbp]
\begin{scriptsize}
\renewcommand{\arraystretch}{1.5}
\begin{center}
\caption{Overview of selected studies using fingerprints-based methods for provenance encoding in IoT networks.\label{tab:comparison_fingerprints}}
\centering
\resizebox{\textwidth}{!}{\begin{tabular}{c | c >{\centering}m{2.5cm} c >{\centering}m{2.5cm} c m{4cm} m{4.3cm}}
\hline
\multirow{2}{*}{\textbf{Reference}} & \multirow{2}{*}{\textbf{Year}}   & \textbf{Provenance Encoding} & \textbf{Provenance Storage} & \multirow{2}{*}{\textbf{Application}} & \textbf{Security} & \multirow{2}{*}{\textbf{Pros}} & \multirow{2}{*}{\textbf{Cons}} \\ 

&  & \textbf{Method} & \textbf{Method} &  & \textbf{Analysis} &  &  \\ 
\hline
\hline
\multicolumn{8}{|c|}{\textbf{Fingerprints}} \\
\hline
~\citet{syed2014} & \refyear{syed2014} & Link Fingerprints & Device Database & Bodyworn Sensors & \xmark & Use of quantization to distill the \gls*{rssi} data into a much smaller size to overcome storage challenges in memory constrained sensor devices. & Considers only single-hop scenarios. Data provenance operates on a per-session basis and is not validated for each packet sent.\\
~\citet{ALAM2014} & \refyear{ALAM2014} & Prime multiplication and Rabin Fingerprints & In-packet buffer & \gls*{iot} sensor network with TelosB motes & \xmark & The use of small number of bits to encode higher number of node IDs and requires fewer packets to construct network-wide provenance. & Provenance information is based on only node IDs, which is not sufficient to encounter many attacks. \\
~\citet{Kamal2018} & \refyear{Kamal2018} & Link Fingerprint & in-packet and server & Multihop \gls*{iot} network usinf MICAz motes & \xmark & The provenance relies on the next hop node using the \gls*{rssi}, which is related to any secrets stored in the nodes. &  The higher the value of correlation coefficient, the higher the percentage of the secured data transfers that can be deceived by an attacker when compromising a group of colliding nodes. \\
\hline
\end{tabular}}
\end{center}
\end{scriptsize}
\end{table}

\medskip

\noindent \textbf{5.8 \SecTab Frameworks using Storing Methods \SecT} Researchers have used various storage methods, using database placement strategies, to develop frameworks designed for diverse \gls*{iot} applications. These methods include a number of techniques for storing and querying provenance information within the database. They offer management solutions to ensure the secure storage of provenance data while meeting security requirements. These frameworks offer a complete cycle for tracking the history of provenance records, covering everything from initial capture to storage and analysis. In this section, we present selected works that have designed such frameworks for managing data provenance \textcolor{black}{and provide an overview of these works in Table~\ref{tab:comparison_storing}.}

\citet{Alkhalil2019} introduce a bio-inspired approach that uses the processes of human thinking to enhance data provenance in \gls*{wsn}. Their proposed Think-and-Share Optimization (TaSO) algorithm modularizes and automates data provenance management in enterprise-deployed \gls*{wsn}. The TaSO algorithm is designed of four phases: Think, Pair, Share, and Evaluate. The authors assess the effectiveness of their TaSO algorithm by evaluating key metrics such as connectivity percentage, closeness to the sink node, coverage, and running time.

\citet{Wang2018} introduce the ProvThings framework, designed for the capture, administration, and analysis of data provenance within \gls*{iot} platforms. Their approach introduces a selective instrumentation algorithm that reduces the collection of provenance data by identifying security-sensitive sources and sinks. This method provides a means to trace complex chains of inter dependencies among \gls*{iot} components. Additionally, the authors created a prototype of ProvThings for the Samsung SmartThings platform and evaluated its effectiveness against 26 \gls*{iot} known attacks. The \gls*{iot} provenance model is based on the W3C PROV-DM~\cite{Groth2013}. The results indicate that ProvThings imposes only a 5\% overhead on physical \gls*{iot} devices while enabling real-time querying of system behaviors. 

Finally, \citet{elkhodr2020} extends previous work that introduced a provenance-based trust management solution to assure data provenance~\cite{Elkhodr2016}. Their Internet of Things Management Platform (IoT-MP)~\cite{Elkhodr2016} establishes trust relationships between communicating \gls*{iot} devices. This work uses the existing capabilities of IoT-MP to enhance privacy protection in \gls*{iot} networks. Furthermore, it introduces a Data Provenance module aimed at enabling the retrieval of data origins and access to device's history, including the networks it has interacted with. The method propose three states for \gls*{iot} devices registered in the IoT-MP platform: New Resident, Visitor, and Returning Resident. Additionally, it includes a database module that is reconstructed to adapt to the changes in the architecture. 

\begin{table}[!htbp]
\begin{scriptsize}
\renewcommand{\arraystretch}{1.5}
\begin{center}
\caption{Overview of selected studies based on frameworks with storing methods for provenance encoding in IoT networks.\label{tab:comparison_storing}}
\centering
\resizebox{\textwidth}{!}{\begin{tabular}{c | c >{\centering}m{2.5cm} c >{\centering}m{2.5cm} c m{4cm} m{4.3cm}}
\hline
\multirow{2}{*}{\textbf{Reference}} & \multirow{2}{*}{\textbf{Year}}   & \textbf{Provenance Encoding} & \textbf{Provenance Storage} & \multirow{2}{*}{\textbf{Application}} & \textbf{Security} & \multirow{2}{*}{\textbf{Pros}} & \multirow{2}{*}{\textbf{Cons}} \\ 

&  & \textbf{Method} & \textbf{Method} &  & \textbf{Analysis} &  &  \\ 
\hline
\hline
\multicolumn{8}{|c|}{\textbf{Frameworks using storing methods}} \\
\hline
~\citet{Alkhalil2019} & \refyear{Alkhalil2019} & Trust model based on fuzzy logic & Network Nodes & \gls*{wsn} & \xmark & Node's trust is based on availability, neighboring nodes evaluation, and message drop rate. & The study needs to consider security issues in-terms of integrity and secure transmission of data. Is not clear how provenance information is encoded and stored. \\
~\citet{Wang2018} & \refyear{Wang2018} & Provenance as sources and sinks & Centralized database & Samsung SmartThings \gls*{iot} Platform & \xmark & Complete platform for provenance tracking in \gls*{iot} applications. Records provenance information by a provenance collector directly from \gls*{iot} devices. & No security mechanism is applied to secure the transmission of provenance information. Multi-hop tracking is not considered. \\
~\citet{elkhodr2020} & \refyear{elkhodr2020} & Store device status & Local database & \gls*{iot} Management Platform & \xmark & Stores the provenance information as device status in a centralized database including the visiting and returning devices. & Provenance information is not secured. Multi-hop model is not taken into consideration. Security requirements in terms of data and provenance are not studied. \\
\hline

\end{tabular}}
\end{center}
\end{scriptsize}
\end{table}

\medskip

\noindent \textbf{5.9 \SecTab Graph-based Provenance \SecT} Graph-based provenance is a method used in data provenance encoding to represent the origin or history of data. These graphs capture the relationships and dependencies among data elements, processes, and transformations, providing a visual or structural representation of how data has evolved over time. Three different types of graph-based provenance representations as descibed below \textcolor{black}{and an overview of these methods is provided in Table~\ref{tab:comparison_graph}.}

\emph{Event-flow graphs} are a type of provenance graph that represents data provenance by capturing events and the flow of data between them. Each event corresponds to a data operation or transformation, and edges in the graph indicate the data flow from one event to another. ~\citet{chang2022} propose a data provenance approach for provenance systems in \gls*{iot} applications based on event-action flows instead of provenance graphs. \textcolor{black}{Event-action} flow is defined as a sequence of events and actions including a time-stamp in an execution trace. They are less complex than provenance graphs due their simple structure form which allows users to understand it easier. In their work, the authors present an event-flow graph to regular users as a static abstraction of every possible provenance graph for \gls*{iot} applications. They dynamically link time-stamped events and actions to statically create event and action nodes. Then, users can query provenance information from the event-flow graph by selecting an event or action node to choose the associated time-stamped actions or events. After users create a query by picking a timestamped event or action, the system will respond with one of two types of information: ``What provenance'' answers questions about which events or actions are triggered by the user's specified action or event, and ``Why provenance'' provides insights into the events or actions responsible for causing the user-specified action or event. The system is developed as a Graphical User Interface (GUI) forming a user-friendly graphical provenance system.

\emph{Provenance graphs} include a variety of graph structures that capture provenance information. These graphs may use different notations and structures to represent data history, depending on the specific needs of the application or research. In a provenance graph, nodes represent data entities, processes, or events, and edges denote relationships, transformations, or dependencies between these nodes. ~\citet{Jaigirdar2023} extend their work, Prov-IoT~\cite{Jaigirdar2020}, to provide security information for provenance graphs. Prov-IoT is a security-aware \gls*{iot} provenance model. In this work, security metadata is integrated with specified security policies within the provenance graphs. They propose an IoT-Health scenario with a number of potential threats: fault packet injection, node cloning, unauthorized access, malicious code injection and denial of service. Three major node types are used to describe the scenario: agent, entity, and activity~\cite{moreau2011}. The terms  Was Associated With (WAsW),  Was Informed By (WInB), Was Derived From (WDeF), and Was Generated By (WGeB) are used to represent the relationships between these three nodes. A general provenance graph for the IoT-Health scenario is generated using the W3C-standardized PROV-DM concept. To evaluate the system and check for potential risks, the approach is evaluated based on six cases: permission violation, missing Web Application Firewall (WAF), intrusion detection, unauthorized access, denial of service, and identifying failed signature verification. 

Finally, the use of \emph{Directed Acyclic Graphs} (DAG) allows directed structures with no cycles. This means that data transformations and dependencies are represented as a directed flow without loops or feedback. DAGs are used to represent provenance information because they are well-structured and provide efficient querying and tracking of provenance records. ProChain framework is a provenance-aware approach of traceability proposed by Al-Rakhami and Al-Mashari~\cite{Al-Rakhami2022} for IoT-based supply chain systems. The IOTA protocol, a third-generation DLT, is used by the ProChain framework. To overcome limitations and provide a scalable, quantum-resistant, and attack-proof solution for the systems built around \gls*{iot}, it makes use of the DAG information structure in contrast to the linear structure used by the blockchain~\cite{Silvano2020}. ProChain enables food item traceability from manufacturing to retailer with the use of several \gls*{iot} sensors and provenance data at each engaged supply chain phase. ProChain strengthens and improves the management and optimization of all operations while also serving as a guarantee for the quality and safety of food. On the Raspberry Pi 3B platform, the ProChain idea is evaluated by simulating an IoT-deployed supply chain. The average measured time and energy consumption were then evaluated to check on the usability of the framework. The authors review and implement the framework to show how it may be used in supply chain systems to couple supply chain data with the IOTA Tangle and generate provenance information by adding data for various payload sizes.

\begin{table}[!htbp]
\begin{scriptsize}
\renewcommand{\arraystretch}{1.5}
\begin{center}
\caption{Overview of selected studies using graph-based techniques for provenance encoding in IoT networks.\label{tab:comparison_graph}}
\centering
\resizebox{\textwidth}{!}{\begin{tabular}{c | c >{\centering}m{2.5cm} c >{\centering}m{2.5cm} c m{4cm} m{4.3cm}}
\hline
\multirow{2}{*}{\textbf{Reference}} & \multirow{2}{*}{\textbf{Year}}   & \textbf{Provenance Encoding} & \textbf{Provenance Storage} & \multirow{2}{*}{\textbf{Application}} & \textbf{Security} & \multirow{2}{*}{\textbf{Pros}} & \multirow{2}{*}{\textbf{Cons}} \\ 

&  & \textbf{Method} & \textbf{Method} &  & \textbf{Analysis} &  &  \\ 
\hline
\hline
\multicolumn{8}{|c|}{\textbf{Graph-based}} \\
\hline
~\citet{chang2022} & \refyear{chang2022} & Event-flow graphs & Provenance server & \gls*{iot} smart apps & \xmark & Observes provenance without using a sophisticated query language by just selecting the appropriate nodes on an event flow graph.  & As the number of events and actions increases, the provenance data becomes large in scale over time, provenance data is not securely stored in the provenance server. \\
~\citet{Jaigirdar2023} & \refyear{Jaigirdar2023} & Provenance graphs &  Cloud & \gls*{iot} Health applications &  \xmark & The model includes security-aware properties at every step of data transmission. Provides the status of each device as data processing mechanism, software running and communication channels properties & Single point of failure where provenance information is stored in the cloud. All the devices/sensors and users need to forward and retrieve this information from it, including auditor, doctor, user, and gateway. \\
~\citet{Al-Rakhami2022} & \refyear{Al-Rakhami2022} & Directed Acyclic Graph & Cloud server & Industrial Internet of Things (IIoT) & \xmark & Scalability, affordability, and quantum robustness are all associated with the adoption of IOTA's distributed ledger technology (DLT). & MQTT protocol is used to store and manage the majority of the data that is gathered by our system, but because it does not execute or impose data encryption, it is not completely safe against tampering. \\
\hline
\end{tabular}}
\end{center}
\end{scriptsize}
\end{table}

\noindent \textbf{5.10 \SecTab Cryptography-based Techniques \SecT} Various cryptographic techniques, including symmetric and asymmetric cryptography methods, hash functions, and digital signatures, are applied to design and implement systems for identifying the source of data and ensuring the integrity of both data and provenance. These concepts are used to propose solutions for data provenance in many applications. \cmmnt{Many researchers propose solutions for data provenance in \gls*{iot} using different cryptographic techniques.} The literature includes the largest number of these solutions when dealing with cryptographic approaches \textcolor{black}{as shown in Table~\ref{tab:comparison_cryptography}.} The trustworthiness of the data items in a sensor network is evaluated by ~\citet{Lim2010}. They compute trust scores for sensor nodes and data packets by using data provenance. These scores provide a method of indicating the level of trustworthiness of both nodes and data items. This method provide security and trustworthiness for sensor networks, yet they fail to address the challenge of retrieving data provenance against various attacks. Moreover, a dictionary-based secure provenance approach for \gls*{wsn} is proposed by ~\citet{wang20162}. They embed path indexes in the provenance instead of the actual data path by using packet path dictionaries. Therefore, compared to the current lossy provenance systems, the compressed provenance size in the proposed lossless approach is smaller. To achieve security requirements such as integrity, availability and authenticity of provenance, the AM-FM sketch method and a robust packet sequence number generating technique are used in the system. Provenance records are encoded at nodes that engage in every stage of data processing and transmission using the suggested dictionary-based system. A secure message authentication code integrates the packet and its provenance together to provide security against any unauthorized change. After verifying the \gls*{macode} during decoding, the BS retrieves the packet's provenance graph. 

While the majority of previous studies have concentrated on how to protect against data manipulation in Home Area Networks (HAN) networks, ~\citet{Chia2017} introduce a security topic that has received less attention in such networks, namely data provenance. To ensure that the stated energy usage is actually consumed and is gathered from the specified node at the exact location, they provide a unique technique based on threshold cryptography and Shamir's secret sharing~\cite{shamir1979}. The authors describe a unique use of secret sharing schemes for home energy monitoring networks, in which a secret, or a single private key, is distributed among all the members who provide data on energy consumption. This system also addresses the issue of location verification. To achieve this, authors incorporate a location generator as an additional component. The location generator employs trilateration techniques based on \gls*{rssi} values and \gls*{rssi} filtering methods to verify the location of the device. This not only ensures that the right power source is under observation but also detects any potential device relocation. Also, ~\citet{suhail2018} provide a solution to the challenge of integrating \gls*{iot} with a system that is aware of data provenance, enabling the tracking of data flow across nodes and monitoring data transformations applied by nodes in the network. They introduce a lightweight method for transmitting provenance information for \gls*{iot} sensor data. This method encodes data provenance information using a hash chain scheme as it traverses each participating node, with the final provenance verification taking place at the destination node. The technique is implemented within the context of the IPv6 Routing Protocol for Low-Power and Lossy Networks (RPL)~\cite{roger2012} in Contiki OS~\cite{dunkels2004}. Each generated data packet includes three fields: packet sequence number, data payload and provenance information (i.e. hash value of node ID). Furthermore, ~\citet{xu2019} present a provenance approach for \gls*{wsn} based on path index differences, where a new packet path is encoded using the index of a highly similar backbone path and the differences between them. Even if the topology of the \gls*{wsn} is unstable, this technique may reach an increased provenance compression rate compared to the dictionary based provenance scheme~\cite{wang20162}. In the presented backbone path selection approach for \gls*{wsn}, the gradient of the node on the data packet path is also determined. In order to cover the commonly used packet transmission channels, the authors also develop a comparable backbone path elimination technique for \gls*{wsn}. To find the packet pathways with the highest similarity in the dictionaries, they develop a locality-sensitive hashing (LSH), SimHash~\cite{Buyrukbilen2014,Charikar2002}, based similarity comparison algorithm. The proposed approach is evaluated according to Total Energy Consumption (TEC) and Average Provenance Size (APS).

Moreover,~\citet{SUHAIL2020} present an approach called Provenance-enabled Packet Path Tracing for \gls*{iot} devices connected through the RPL protocol. Their scheme involves including sequence numbers into the routing entries of the forwarding nodes' routing table, establishing a node-level provenance. Additionally, they introduced a system-level provenance that encompassed destination and source node IDs, enabling complete packet trace capture. To retrieve the entire data provenance using this approach, it is essential to sequentially access the storage space of each node along the routing path. Hence, the base station cannot independently decompose the complete provenance of each data packet. Liu and Wu~\cite{liu2020} introduced an algorithm for compressing provenance called index-based provenance compression. To reduce the overall size of the provenance data, their approach combines the concept of typical substring matching with path identifier and path index to represent path information within data provenance. Additionally, they expand the data provenance scheme to include attack detection and present a method for identifying malicious nodes based on this expanded scheme. The proposed scheme falls short in terms of ensuring data integrity, lacks a thorough security analysis within a defined threat model and results in computationally intensive operations. Additionally, Tang and Keoh~\cite{tang2020} present a methodology to ensure the unaltered reporting of energy usage data of home appliances. The proposed approach guarantees the collection of data from the correct and trusted source at the specified location. This framework is specifically designed for Home Area Networks (HAN) within a smart metering infrastructure, with the primary objective of confirming the authenticity of data, source identity, and location. The \gls*{macode} is used to verify data authenticity and integrity. It should be noted that the authors assume the trustworthiness of the receiver (i.e. smart plug), while acknowledging the possible vulnerability of one of the two senders (i.e. smart plug or magnetic sensor) to attacks. To achieve location authenticity verification, the system integrates a third sender (i.e. a Bluetooth device), which operates independently and does not collude with other system components.

In their research, ~\citet{Xu2022} introduce two provenance schemes for Wireless Sensor Networks. The first is the Path Index Differences-based Provenance (PIDP) scheme, where provenance information is encoded using differences in packet path indices. Specifically, it links an index representing the packet path's similarity to the main backbone path and the variation between the actual packet path and this backbone path. Additionally, the authors propose a second scheme known as the Path Hash Value-based Provenance (PHP) scheme. In this approach, provenance data is encoded as a combination of the data source node's ID and a part of the hash value taken from the packet's path. Both schemes are evaluated based on their provenance compression rate and energy conservation rate, and these metrics are then compared to those of the DP scheme. Xu and Wang~\cite{xu20222} introduce a provenance scheme called Multi granularity Graphs-based Stepwise Refinement Provenance (MSRP). In this scheme, they use mutual information between pairs of nodes as a similarity index to categorize node IDs. This categorization forms the basis for generating multi granularity topology graphs. Additionally, they apply the Dictionary-based Provenance (DP) scheme for stepwise encoding of the provenance information. The Base Station (BS) follows the same stepwise approach for provenance recovery and simultaneously conducts data trustworthiness evaluation during decoding. The performance of the MSRP scheme is thoroughly evaluated through a combination of simulations and testbed experiments. Results show that the scheme achieve high provenance compression rate, energy usage and efficiency of data trustworthiness assessment. 

\begin{table}[!htbp]
\begin{scriptsize}
\renewcommand{\arraystretch}{1.5}
\begin{center}
\caption{Overview of selected studies using cryptography-based methods for provenance encoding in IoT networks.\label{tab:comparison_cryptography}}
\centering
\resizebox{\textwidth}{!}{\begin{tabular}{c | c >{\centering}m{2.5cm} c >{\centering}m{2.5cm} c m{4cm} m{4.3cm}}

\hline
\multirow{2}{*}{\textbf{Reference}} & \multirow{2}{*}{\textbf{Year}}   & \textbf{Provenance Encoding} & \textbf{Provenance Storage} & \multirow{2}{*}{\textbf{Application}} & \textbf{Security} & \multirow{2}{*}{\textbf{Pros}} & \multirow{2}{*}{\textbf{Cons}} \\ 

&  & \textbf{Method} & \textbf{Method} &  & \textbf{Analysis} &  &  \\ 
\hline
\hline
\multicolumn{8}{|c|}{\textbf{Cryptography-based}} \\
\hline
~\citet{Lim2010} & \refyear{Lim2010} & Trust scores & N/A & Sensor Network & \xmark & Practical solution for trustworthiness assessment. & The method is based on the principle that the more trustworthy data a source provides, the more trusted the source is considered. Many attacks can deceive the system and overcome the trustworthiness of data. \\
~\citet{wang20162} & \refyear{wang20162} & Path Index and \gls*{macode} & Distributed database & \gls*{wsn} & \usym{1F5F8} & Embeds path index instead of data path, which reduces the size of the embedded information in each packet. Each packet path is stored at the forwarding nodes in the network. & Provenance information is only accessible at the BS and cannot be verified at each stage of the path, provenance data is not securely transmitted and stored. \\
~\citet{Chia2017} & \refyear{Chia2017} & Shamir secret sharing and threshold cryptography & Smart meter & Home Energy Monitoring Networks & \xmark & Achieves source identity authenticity, location authenticity, data consistency and source data authenticity.  & The proposed solution does not consider the storage of provenance information, multi-hop architecture and provenance encoding. \\
~\citet{suhail2018} & \refyear{suhail2018} & Hash chain & In-packet & Sensor network & \xmark & Keeps track of data packets using a chain of provenance records that store that hash of traversed node ID. & Integrity is verified at the destination. Provenance size grows very fast as the number of forwarding nodes increase, link overhead due size increase in forwarded packets. \\
~\citet{xu2019} & \refyear{xu2019} & Path index differences & Network nodes & \gls*{wsn} & \xmark & High provenance compression rate. Whenever the number of hops increases the provenance size nearly remains at a constant level. & Provenance information is based on a few number of data features. Dictionaries at each node increase in size as the network scale and the data packets increase.\\
~\citet{SUHAIL2020} & \refyear{SUHAIL2020} & In-packet embedding & Routing table & RPL-based \gls*{iot} network & \usym{1F5F8} & Constant provenance size, used energy consumption, enhanced provenance generation time. & Considers robustness against only three attacks. Provenance information is node ID and sequence number.  \\
~\citet{liu2020} & \refyear{liu2020} & Common substring matching & Distributed database & Multihop \gls*{iot} network & \xmark & Malicious node identification. High provenance decoding accuracy. Stable provenance length after all the path have been traversed & Many provenance fields that increase the size compared to other techniques. Lack of security analysis against different types of attacks.  \\
~\citet{tang2020} & \refyear{tang2020} & \gls*{macode} & in-packet & HAN with smart metering & \usym{1F5F8} & Use of a symmetric key approach to improve efficiency over asymmetric key-based approaches. & Needs to integrate a third sender to the system to achieve location verification. Solution for single-hop data transfer scenario only. \\
~\citet{Xu2022} & \refyear{Xu2022} & Path index and Packet path hash value & Distributed node database & Zigbee \gls*{wsn} & \xmark & Reduces the negative impact of network topology changes. Maintains high provenance compression and keeps the provenance size constant & Increase in provenance size as network becomes larger in the PIDP scheme. Lacks analysis to different security threats. \\
~\citet{xu20222} & \refyear{xu20222} & Dictionary-based provenance scheme/ Hash functions  & Network nodes & ZigBee sensor network & \xmark & \gls*{wsn} topology graph is presented as a series of different granularity topology graphs. Encoding provenance on high granularity levels which skips provenance updating at some nodes. Decrease in the provenance updating latency. & The scheme yields high cost in terms of computation and storage in sensor nodes, which are resource constrained and require lightweight schemes. \\

\hline
\end{tabular}}
\end{center}
\end{scriptsize}
\end{table}

\medskip

\noindent \textbf{5.11 \SecTab Physical Unclonable Functions \SecT} Physical Unclonable Functions (\gls*{pufs}) provide a hardware-oriented system that generates a response for a specific challenge, ensuring uniqueness for each device in the system. \gls*{pufs} facilitate the authentication of \gls*{iot} devices that generate sensitive data while maintaining device anonymity. Through the presence of a trusted third-party verifier, the identities of these devices can be verified without compromising their anonymity~\cite{arun2013}. \textcolor{black}{The selected techniques are shown in Table~\ref{tab:comparison_puf}.} \citet{aman2019} developed two secure protocols for data provenance in \gls*{iot} networks, aiming to achieve authentication and privacy preservation. These protocols address two different scenarios: the first scenario involves a direct connection between an \gls*{iot} device and a wireless gateway, while the second scenario deals with \gls*{iot} devices indirectly connected to the wireless gateway through multiple hops of other \gls*{iot} devices. Both protocols use \gls*{pufs} in addition to wireless link fingerprints generated from the \gls*{rssi} between communicating nodes. Through experimental analysis, the authors show that these protocols achieve high efficiency in terms of computational complexity and energy usage, while also obtaining robustness against various physical and cloning attacks. \citet{aman2021} propose an analytical model to create a mechanism that enables the establishment of data provenance in \gls*{iot} systems. Their approach incorporates \gls*{pufs} and the extraction of fingerprints from the wireless channel, along with the implementation of mutual authentication and anonymity measures, all aimed at achieving robust data provenance. The approach lacks consideration for the multi-hop scenario and fails to adequately address tracking of data packet provenance. Moreover, Hamadeh and Tyagi~\cite{Hamadeh2021} propose an approach which combine two solutions: data provenance and privacy prserving in \gls*{iot} networks. They provide trustworthiness and dependable \gls*{iot} networks by using \gls*{pufs} with non-interactive zero knowledge proof. In this method, an \gls*{iot} device has the capability to transmit data to its respective server without revealing its identity, as it provides proof of ownership. In particular, the method under consideration is designed to validate that the authorized device executed an authorized program for creating or modifying data. Authors introduce a privacy-centric data provenance protocol. To validate practicality and effieciency, they developed the protocol using Altera Quartus and subsequently implemented it on an Altera Cyclone IV FPGA. 

\begin{table}[!htbp]
\begin{scriptsize}
\renewcommand{\arraystretch}{1.5}
\begin{center}
\caption{Overview of selected studies using watermarking and Physical Unclonable Functions  for provenance encoding in IoT networks.\label{tab:comparison_puf}}
\centering
\resizebox{\textwidth}{!}{\begin{tabular}{c | c >{\centering}m{2.5cm} c >{\centering}m{2.5cm} c m{4cm} m{4.3cm}}
\hline
\multirow{2}{*}{\textbf{Reference}} & \multirow{2}{*}{\textbf{Year}}   & \textbf{Provenance Encoding} & \textbf{Provenance Storage} & \multirow{2}{*}{\textbf{Application}} & \textbf{Security} & \multirow{2}{*}{\textbf{Pros}} & \multirow{2}{*}{\textbf{Cons}} \\ 

&  & \textbf{Method} & \textbf{Method} &  & \textbf{Analysis} &  &  \\ 
\hline
\hline
\multicolumn{8}{|c|}{\textbf{Physical Unclonable Functions}} \\
\hline
~\citet{aman2019} & \refyear{aman2019} & \gls*{pufs} and \gls*{rssi} & Device memory and server & Indoor laboratory \gls*{iot} environment & \usym{1F5F8} & Unclonability and robustness against physical attacks through avoiding the need to store secret keys. & Provenance information is based only on device’s pseudonym identity and \gls*{rssi}, which is not enough to obtain any attack attempt in packet drop, replay, and modification. \\
~\citet{aman2021} & \refyear{aman2021} & \gls*{pufs}, wireless fingerprints  & server database & \gls*{iot} network with MICA-Z motes & \usym{1F5F8} & \gls*{iot} devices do not store secrets in their memory. Privacy preservation. Resilience against Ephemeral Secret Leakage (ESL) attacks. & Requires computation of many session keys for each node. Lacks security analysis against different type of attacks.  \\
~\citet{Hamadeh2021} & \refyear{Hamadeh2021} & \gls*{pufs} & Network nodes and server & FPGA Altera Cyclone & \xmark & Source identity authenticity through PUF. \gls*{iot} node is able to send anonymously  data to the server. & The system does not consider the traceability of data packets along the data path. Multi-hop model with the presence of forwarding devices is not considered. Selecting and computing a secret key in each round results in computational overhead. \\
\hline
\end{tabular}}
\end{center}
\end{scriptsize}
\end{table}

\medskip

\noindent \textbf{5.12 \SecTab Blockchain-based Solutions \SecT} Distributed ledger technologies, like blockchain, have gained visibility for maintaining the security and privacy of provenance data. These blockchain platforms serve as the primary means to ensure data integrity by creating transactions as provenance records that are chained in a secure manner. Once recorded on the blockchain, these records cannot be altered or removed, thereby establishing a trustworthy system for verifying the integrity of the provenance information in a network environment that is vulnerable to different attacks. \textcolor{black}{Table~\ref{tab:comparison_blockchain} provides an overview of the selected studies that uses blockchain-based methods to maintain the security of provenance data in \gls*{iot} networks.}

\citet{Nathalie2017} propose a framework in \gls*{iot} environments to maintain and secure \gls*{iot} provenance data, whose main features are: (i) use of a completely distributed lightweight keyless blockchain element (i.e. Keyless Signature Infrastructure Module (KSI)~\cite{ahto2013,ahto2014}) to guarantee that the integrity of the provenance data is protected; (ii)  maintenance of provenance information confidentiality by granting specific access controls to various parties as required; and (iii) high level of availability of provenance. In addition, provenance data in \cite{Nathalie2017} can be subject to restricted access policies. The integration of cryptographic algorithms in the solution ensures confidentiality since it is difficult to manage the flow of data and its provenance in \gls*{iot} contexts. 

\citet{zeng2018} propose a blockchain-based data provenance scheme (BCP) which deploys a distributed blockchain database that is connected to the sensor network through edge computing. In this scheme, each node updates the provenance records along the packet data path. These records are obtained by high performance nodes (H-nodes), which are edge computing nodes placed either above or close to the \gls*{wsn}, through the process of packet sniffing~\cite{shi2016}. Then, the base station retrieves the provenance records by querying the H-nodes. Moreover, provenance records are stored in a blockchain-chain database after encryption through invoking a smart contract running on Ethereum Virtual Machine (EVM)~\cite{wood2014}. In this model, each provenance record contains node ID, sequence number, hop count, sequence numbers of the aggregated packets, and the number of times a packet is aggregated.

In their work, ~\citet{Javaid2018} propose a solution for data integrity and data provenance in \gls*{iot} networks by using \gls*{pufs} and a blockchain variant with smart contracts that is Ethereum. This method is called BlockPro. \gls*{pufs} are used to maintain data provenance by providing unique hardware fingerprints. To overcome data tampering attacks and block unregistered devices, Ethereum is used as a decentralized digital ledger. With the presence of expanding sequence of records, data undergoes initial validation before being permanently stored on the blockchain. Once stored, it can not be tampered with or modified, thus ensuring data integrity. ~\citet{Sigwart2020} implement an \gls*{iot} data provenance framework based on smart contracts using a generic data model to provide data provenance capturing, storing, and querying functionalities for different \gls*{iot} use cases. This work extends their approach proposed in~\cite{sigwart2019}. The framework uses the data provenance model by ~\citet{Olufowobi2017}. The authors conduct an assessment of the proposed framework with specified requirements through the implementation of a proof-of-concept using Ethereum smart contracts. In this approach, a provenance record ($prov(dp)$) for a data point ($dp$) is composed of a 3-part structure. It links the address ($addr(dp)$) of $dp$ (i.e. essentially an identifier ID) with the set of provenance records associated with the data points that are used to create a $dp$ (referred to as $inputs(dp)$), and includes a context element ($context(dp)$). This element includes information for provenance purposes, such as details about the agents involved in the computation of the data point, timestamp, location, or the specific execution context within the \gls*{iot} system. 

According to ~\citet{liu20202}, a multilayer provenance query index and a blockchain-based architecture are proposed for the network provenance in the \gls*{iot}: blockchain-based secure and efficient distributed network provenance (SEDNP). For effective representations in the Verifiable Computation (VC) framework, the design integrates range, keyword, and K-hop ancestor queries into a single model. The authors also create a digest hashing method that verifies the provenance log and index. With constant-size digests, they decrease the storage and processing costs associated with on-chain transactions regardless of the volume of data. Along with extensive security research, they formalize preserving security to record the requirements for the validity and integrity of the query results. Finally, using a proof-of-concept approach that combines the Pinocchio VC framework and a testing blockchain network, the authors analyze the implementation issues. For cloud-based \gls*{iot} networks, ~\citet{Siddiqui2020} provide an application layer data provenance system that relies on an execute-order architecture. By using outsourced encryption on Edge nodes using Ciphertext-Policy Attribute-based Encryption (CP-ABE), it allows fast transaction throughput on the blockchain network with minimal security overhead. The workload on the \gls*{iot} nodes is reduced since every communication between \gls*{iot} devices is connected to a blockchain network and recorded on authorized blockchain peers. Smart contracts are used to store the provenance data on the blockchain. The \gls*{iot} Device Registration smart contract begins when an \gls*{iot} node connects to the network. 
The Data Transfer smart contract is run whenever a message is transmitted from one \gls*{iot} node to another. This is responsible for preserving the provenance information on the blockchain. The Provenance Verification smart contract, which is in charge of confirming the provenance data, is then run when it has been confirmed that the message was transmitted by the appropriate \gls*{iot} node. 

In their work, Porkodi and Kesavaraja~\cite{Porkodi2021} present a framework to secure the data provenance by introducing blockchain and access control policies in \gls*{iot} systems. They use hybrid attribute-based encryption to securely transmit provenance information. The proposed method is evaluated using different performance metrics such as computational cost and the throughput of encryption/decryption, and the key strength is calculated using the effect of avalanche. Additionally, the authors conducted experiments to prove that the proposed approach reduces computational cost and achieves high throughput. 

Another solution is presented by~\citet{yin2022}, which use blockchain and smart contract to develop a data provenance scheme for \gls*{iot} applications. The scheme includes a smart contract on the blockchain with reasonable access control policy. The access authorization for users and data sources is limited to maintain security of generated data. For data security and data integrity, the suggested approach implements two data structures: provenance record and provenance record set. To achieve data provenance, \cmmnt{use} blockchain's non-repudiation is used. Access to the data is made secure by this structure. The provenance system can continue to function and the smart contract may impose the data owner's policy. To evaluate the feasibility of the proposed scheme, Ethereum network test is used for verification. Additionally, ~\citet{sun2022} construct a blockchain-based \gls*{iot} data provenance model through adapting the PROV data model (PROV-DM)~\cite{missier2013} to address the issue of recording provenance data generated by a multi-layer \gls*{iot} system. The provenance elements are defined by PROV-DM along with their connections. The W3C provenance family of standards has this conceptual data model as its basis. To create an \gls*{iot} data provenance model, it is therefore required to improve and expand the PROV-DM. In this work, authors present the needed requirements  to design and build a robust data provenance model. Then, they propose a model based on PROV-DM using a blockchain network to secure provenance records from being tampered. To demonstrate how the suggested approach may be used to build a provenance graph and identify which agent is operating \gls*{iot} data abnormally, they provide an application scenario in the \gls*{iot} trustworthy data sharing system.

ProvNet, a distributed data sharing system that may ensure data ownership and accurately record and preserve data sharing provenance, is proposed by~\citet{chenli2022}. \cmmnt{They examine two cases:}
Users can share data in two ways: through the same service provider, where provenance verification and storage will be handled by the service provider and certain users; or through other service providers, who will work together to maintain a provenance graph and authenticate sharing records. They suggest a blockchain variant structure that can provide both forward and backward tracking in order to store the provenance information during the sharing process. A directed graph is appropriate for keeping the provenance records because of the nature of data sharing, which allows senders to re-share the datasets they receive and allows datasets to be shared among many recipients simultaneously. ProvNet suggests storing the provenance records in a networked blockchain, or blocknet, as an alternative to a single blockchain. In order to perform forward tracking, ProvNet selects redactable blocks~\cite{Ateniese2017,Christina2009} and Chameleon Hash~\cite{chameleon1998,Jan2017}, allowing a block to record the hash value of its subsequent block while maintaining its own hash value.

\begin{table}[!t]
\begin{scriptsize}
\renewcommand{\arraystretch}{1.5}
\begin{center}
\caption{Overview of selected studies using blockchain-based methods for provenance encoding in IoT networks.\label{tab:comparison_blockchain}}
\centering
\resizebox{\textwidth}{!}{\begin{tabular}{c | c >{\centering}m{2.5cm} c >{\centering}m{2.5cm} c m{4cm} m{4.3cm}}
\hline
\multirow{2}{*}{\textbf{Reference}} & \multirow{2}{*}{\textbf{Year}}   & \textbf{Provenance Encoding} & \textbf{Provenance Storage} & \multirow{2}{*}{\textbf{Application}} & \textbf{Security} & \multirow{2}{*}{\textbf{Pros}} & \multirow{2}{*}{\textbf{Cons}} \\ 

&  & \textbf{Method} & \textbf{Method} &  & \textbf{Analysis} &  &  \\ 
\hline
\hline
\multicolumn{8}{|c|}{\textbf{Blockchain-based}} \\
\hline

~\citet{Nathalie2017} & \refyear{Nathalie2017} & Keyless Signature Infrastructure Module (KSI) & Blockchain & N/A & \xmark & Provenance data that is secured can only be accessed by authorized users. Lightweight and scalable architecture for \gls*{iot} applications. & Data points generated from \gls*{iot} devices/sensors are not secured when communicated to the gateway and then to the policy engine and KSI before storing the provenance data as a transaction in the blockchain. \\
~\citet{zeng2018} & \refyear{zeng2018} & Ethereum blockchain using edge computing nodes & blockchain database & Raspberry Pi Nodes and micaz motes & \xmark & No provenance compression is needed. Secure provenance storage through blockchain database. & Each data packet requires a transaction for updating provenance information. Large number of generated data packets from sensor nodes, requiring a complex and costly method to store and query each provenance record for the data packets from the blockchain. \\
~\citet{Javaid2018} & \refyear{Javaid2018} & \gls*{pufs} & Blockchain & Linux working environment & \usym{1F5F8} & Prone to single point of failure due to the decentralized architecture. Smart contracts enable a safe and secure mechanism for the transmission, authentication and storage of requests. & Each PUF Challenge Response Pair (CRP) and the address of \gls*{iot} device is stored by the smart contract. The huge number of data generated makes it complex to store this amount of data using a blockchain. \\
~\citet{Sigwart2020} & \refyear{Sigwart2020} & Data points & Blockchain-based & General & \usym{1F5F8} & A general framework for different \gls*{iot} applications. Layered architecture of smart contracts. & No consideration for the secure transmission of provenance records. New architecture is needed in the \gls*{iot} platform. \\
~\citet{liu20202} & \refyear{liu20202} & Hash function and K-Hop Ancestor & Blockchain & \gls*{iot} application with blockchain network & \usym{1F5F8} & Examines existing security concerns in the distributed \gls*{iot} network architecture. Uses blockchain as the fundamental architecture for storing and retrieving cross-domain provenance data by using its decentralization and immutability. & Querying provenance information needs to be optimized due to the cost of retrieving on/off blockchain storage. \\
~\citet{Siddiqui2020} & \refyear{Siddiqui2020} & Ciphertext-Policy Attribute based Encryption (CP-ABE) & Blockchain and centralized database & Cloud based \gls*{iot} & \xmark & By applying partial signatures, it is possible to offload the blockchain method and associated overhead from the \gls*{iot} node to the edge nodes. & Each provenance record must be stored twice (requiring additional storage and communication overhead), with the block being saved in the provenance database and the provenance data being published on the blockchain network by a provenance auditor. \\
~\citet{Porkodi2021} & \refyear{Porkodi2021} & Hash function and symmetric encryption & Blockchain & \gls*{iot} network & \usym{1F5F8} & Multiple levels of authorities. Lightweight key management. & Verifying data origin is not satisfied. Needs for provenance storage. Provenance information is not sufficient to establish trustworthiness in the system. \\
~\citet{yin2022} & \refyear{yin2022} & Smart contract & Blockchain & Ethereum network & \xmark & Users can not access provenance information without access permission from data owner. Each time an operation is performed the authority is checked. & Provenance information needs to be communicated securely with the blockchain storage. Each operation need to be set as a transaction to be stored within the blockchain, which makes it complex. \\
~\citet{sun2022} & \refyear{sun2022} & Hash function and Homomorphic Signature & Blockchain & Multi-layer \gls*{iot} applications & \xmark & Detects abnormal data operations using the provenance graph with integrity verification using a signature and a hash function on each \gls*{iot} node. & Provenance information is not securely communicated with the edge node or gateway. \\
~\citet{chenli2022} & \refyear{chenli2022} & Directed graph and Chameleon Hash & Blockchain & Decentralized data sharing applications & \usym{1F5F8} & Performs both forward and backward tracking. Uses networked blockchain instead of single blockchain for storing provenance records. & Costly in terms of computation and storage for resource limited devices such as sensors and devices in \gls*{iot} networks. \\

\hline
\end{tabular}}
\end{center}
\end{scriptsize}
\end{table}

After analyzing the selected state-of-the-art papers based on our proposed taxonomy in Figure~\ref{fig:taxonomy}, \textcolor{black}{the findings of this analysis are shown in Tables~\ref{tab:comparison_cryptography} to ~\ref{tab:comparison_storing} and Tables~\ref{tab:category} to ~\ref{tab:attacks} in Appendix~\ref{appendix}.} Each paper is analyzed based on the provenance encoding method, provenance storage method, application, security analysis, advantages and shortcomings. The findings from this in-depth analysis of the presented literature is shown in details in the Discussion section below.

\section{Discussion}
\label{sec:discussion}
Data provenance can be used to detect errors in the different stages of data generation and processing enabling the system to detect the nodes that produced those errors~\cite{Yu2017,yan2017,YU2019}. In addition, storing detailed information about data in the provenance record allows for data recovery when data is no more usable to ensure availability and achieve normal data communication within different network entities~\cite{foster2003}. Data provenance enhances data readability when it includes detailed information about the data's origin and processing~\cite{jagadish2004}. Furthermore, data provenance enhances data clarity, ensure data reliability, and facilitates data reuse~\cite{jensen2013}. One of the most important features of data provenance is providing the system with the ability to asses the trustworthiness of generated data through different secure provenance techniques~\cite{sabaa2007}.  However, the size of provenance records depends on the number of nodes involved in generating the provenance information and the quantity of attributes to be included in each record. To obtain a provenance chain that satisfies security requirements, it is essential to include many attributes that describes the origin, transformation, data path, data quality and any modifications the data has undergone. In large-scale networks with an increasing number of nodes, the size of provenance information grows rapidly, posing significant challenges in terms of storing and querying these records. This can limit the efficiency of provenance analysis. There should be a trade-off between the number of attributes included in the records and the limitations in the computational capabilities of the system. This requires to determine the most important security requirements of the system and the needed attributes that satisfies it with minimum storage and querying overhead.

The presence of constrained network components and limited resources in \gls*{iot} environments presents several challenges for data provenance schemes. Based on the analyzed literature, any data provenance scheme designed for such environments must address multiple challenges to be effective~\cite{salmin2013, ALKHALIL2017, suhail2016}. One of the important challenges is minimizing bandwidth consumption while ensuring high data processing and throughput in the provenance infrastructure. Properly indexing the provenance records is also essential due to the extensive nature of complete provenance. Queries often involve more complex operations than simple name-based databases, requiring users to search for specific data sets based on subsets of attributes and values within the provenance chain. Different users may query different attributes depending on their objectives, necessitating efficient indexing structures for databases across various dimensions. Additionally, managing the size of provenance data efficiently is crucial, as provenance records in large-scale systems typically exceed the size of the actual data when processed and transmitted.

Establishing secure transmission of provenance information and enabling the detection of malicious attacks with fast response from provenance management systems is vital. Efficient storage of provenance data is equally important, particularly as original data undergoes multiple hops and accumulates complex processing histories, resulting in large provenance information. There should also be flexibility in querying provenance data to allow reconstruction when an authorized entity queries it from the provenance storage entity. Collecting provenance information poses its own challenges, as different data features may be collected and stored based on the service and requirements of the IoT system. Often, the system requires the collection of various operations or information about forwarding entities, necessitating effective handling of these records from the generation of data to its final receiving node.

\textcolor{black}{Based on the insights gained from the analysis and in-depth study of the security techniques for data provenance presented in Section~\ref{sec:techniques}, we can better understand the critical challenges and considerations outlined above. These security techniques not only address the issues of trust and integrity within data provenance but also inform our recommendations for enhancing the development and implementation of effective provenance solutions in IoT networks.}

\textcolor{black}{In Table~\ref{tab:category}, we provide details on the evaluation mechanisms used by the various techniques, categorizing them as hardware-based, cloud-based, or simulation-based (software-based or simulator). This resource aims to assist researchers in exploring the different implementation tools and simulation software associated with each provenance category. Our analysis reveals that the majority of studies predominantly use simulation-based approaches to implement and test their techniques. These software-based simulators offer flexibility and cost-effectiveness, making them a popular choice among researchers. However, only a limited number of studies combine both hardware and simulator implementations, which could provide a more thorough evaluation of the techniques in real-world scenarios. This observation highlights a potential gap in the literature, where practical, hardware-based testing could validate the effectiveness of proposed solutions under more realistic conditions. Encouraging a balance between simulation and hardware implementations may lead to more robust and applicable techniques for data provenance in IoT environments.}

\textcolor{black}{Moreover, in Table~\ref{tab:performance}, we present the performance metrics evaluated by various security techniques for data provenance in IoT, based on a selection of recent literature. Our analysis reveals that while several metrics are available for assessing the effectiveness of provenance mechanisms—such as provenance length, energy consumption, data packet size, link-loss rate, detection rate, false positive rate, false negative rate, and computation time—most existing studies focus primarily on a limited subset of these, typically provenance length, energy consumption, and computation time. This selective focus indicates that while these metrics are essential, there is limited research that thoroughly evaluates techniques across multiple performance dimensions. For instance, metrics like detection rate, false positive rate, and false negative rate are important for assessing the security and reliability of provenance techniques. However, few studies include these alongside resource-related metrics, such as energy consumption and computation time. Similarly, link-loss rate, which can impact communication stability in IoT networks, is often overlooked. By examining Table~\ref{tab:performance}, we observe a need for more holistic evaluations that consider the full spectrum of performance metrics. Expanding future work to address this broader set of metrics could provide more robust assessments of the practical feasibility of these techniques. This would also align with IoT requirements, where both resource efficiency and high security are critical. The lack of studies that comprehensively address multiple performance metrics highlights an area for further research in developing balanced, resource-conscious, and security-oriented solutions for IoT provenance.}

\section{Research challenges and open issues}
\label{sec:challenges}

\textcolor{black}{Following our analysis of the security techniques presented in Section~\ref{sec:techniques} and the discussions in Section~\ref{sec:discussion}, we aim to foster further dialogue and propose recommendations for advancing data provenance in IoT networks by outlining several open problems, research gaps, challenges, and limitations identified in the reviewed literature.} 

\noindent \textbf{9.1 \SecTab Research challenges \SecT} The integration of data provenance with \gls*{iot} raises critical security concerns and in this section we summarize some of the important challenges observed while integrating provenance with \gls*{iot}.

\begin{itemize}
\item A first challenge deals with \emph{provenance records processing and storage}. Indeed, provenance information may be larger than the data itself, since provenance records gets larger as the number of forwarding nodes increase. While some applications may involve a small-scale network, there is a need for a complete provenance representation in the provenance chain. This representation requires a number of attributes that need to be stored in the provenance records which is also an issue with storage. Additionally, the diversity of \gls*{iot} devices and sensors leads to a variety of data types and formats. 

\item Provenance records must also accommodate to the different types, leading to a larger and more complex datasets. Also, to provide useful information from the history and processing of data, provenance records need to be detailed and granular. This granularity adds to the size of the provenance dataset, especially when capturing detailed information about each data transformation. In large-scale \gls*{iot} deployments, devices are often interconnected in complex networks. Provenance records need to traverse these networks, leading to additional metadata and size considerations as data moves across multiple devices and systems. Moreover, \gls*{iot} devices may have limited storage and bandwidth capacities. Transmitting, storing, and managing large provenance datasets can load these resources, impacting the overall performance and efficiency of \gls*{iot} networks. There exist some solutions to overcome this problem such as compression techniques. These techniques have high loss rate and increase the computational complexity of the system with limited resources. It is challenging to take into account the required information to be stored in provenance records and maintain, at the same time, processing and storage overheads.
Addressing the challenge of provenance size in \gls*{iot} networks requires careful consideration of storage solutions, data compression techniques, and protocols for optimizing data transfer and processing. Balancing the need for detailed provenance information with the constraints of \gls*{iot} environments is very important for effective and efficient provenance management in \gls*{iot} networks.

\item A second challenge deals with \emph{provenance attachment to data packets}.  Making sure provenance flows with data is a challenging task. Provenance is a type of metadata which increases in size as the number of forwarding nodes increases. In an \gls*{iot} network, data often traverses diverse and resource-constrained devices, making it essential to track the origin, transformations, and actions performed on the data. Embedding provenance information directly into data packets can be challenging due to constraints such as limited bandwidth, energy, and processing capabilities of \gls*{iot} devices. Maintaining a balance between the requirements of data provenance and the limitations of \gls*{iot} networks is essential. Addressing this issue involves developing efficient and lightweight methods for attaching and transmitting provenance records with data packets and ensuring that the provenance information is captured throughout the data path across the \gls*{iot} network without causing significant overhead or affecting the functionality of the devices.

\item Two additional challenges are related to \emph{provenance collection} and \emph{provenance privacy}. The former, provenance collection, deals with the large amounts of data generated, usually in real time, in \gls*{iot} network. This requires the need for tracking its origin, transformations, and actions resulting in an overhead for collecting provenance records. Additionally, the heterogeneity of \gls*{iot} devices introduces complexities in standardizing provenance formats, as different devices may generate diverse data types and use different communication protocols. The resource-constrained nature of many \gls*{iot} devices makes provenance collection more difficult, as it requires \cmmnt{usage of} energy consumption, storage limitations, and processing capabilities. Furthermore, extracting the provenance information from the networks that are designed without considering the possibility of need for querying provenance information is a challenging task. 

\item Provenance privacy emerges as a challenge in \gls*{iot} networks due to the sensitive nature of data and the wide number of different interconnected devices. Provenance records, which trace the origin and transformations of data, provide details about the context and usage history of information. In an \gls*{iot} system, where diverse and personal data is continuously generated, maintaining the privacy of individuals becomes essential. The challenge is to balance the need for important provenance information while protecting user privacy. Dealing with these issues requires careful consideration of secure transmission, encryption protocols, secure storage and access controls to ensure that while provenance records remain effective for achieving security requirements, they do not compromise the privacy rights of individuals and entities whose data contributes to the \gls*{iot} network. Developing robust privacy-preserving mechanisms becomes essential to address these concerns and achieve trust in the secure development of \gls*{iot} networks.

\end{itemize}

\noindent \textbf{9.2 \SecTab Open issues \SecT} First, we identify  some representative open issues and potential research directions which can help in exploring different aspects of data provenance integration in \gls*{iot} networks.

\begin{itemize}
    \item \textbf{\textit{Privacy and security of data stored on the blockchain:}} Integrating a blockchain with the \gls*{iot} network provides a robust database that provides integrity to the provenance records stored as transactions. These records include sensitive information about the creation, processing and transmission of the data packets. Hence, it is essential to preserve privacy while maintaining the integrity of the complete provenance chain. Attackers may try to obtain secret information by analyzing the provenance chain which the blockchain holds. It is important to secure the records stored on the blockchain. 
    \item \textbf{\textit{Range of studied attacks:}} Most of the proposed approaches focus on a very few and limited number of attacks. In many cases, researchers take into account either data attacks or provenance attacks. Attackers aim to deceive the system by targeting both data and provenance records. The majority of existing works focus on the forgery and modification attacks, which are the most common threats. It is clear, from Table~\ref{tab:attacks}, that almost all the proposed methods do not consider chain tampering. As mentioned above, a complete provenance information of a particular data packet is created by connecting provenance records in a chain. The objective of an attacker is to change this chain's sequence and contents to affect the accuracy and dependability of the provenance data. Hence, to develop any solution for provenance in \gls*{iot}, \cmmnt{it should take into account} the robustness of the system against data and provenance attacks must be taken into account.
    \item \textbf{\textit{Watermarking as a solution:}} Maintaining data integrity through procedural requirements is the major goal of watermarking system design. These requirements cover node ownership, data integrity, and bandwidth establishment. Embedding a watermark with data is the main technique to achieve the  requirements in watermarking schemes~\cite{salmin2013,Salmin2015,dai2008,hussain2014,panah2015}. The criteria used in the generation of watermarking schemes is not restricted to \gls*{wsn}; it also includes the security of multimedia programs~\cite{Sultana2011,hasan2009,prasad2010,mehdi2015} and database formation~\cite{jain2015,gosavi2015,khan2013}. This wide range of domains allows watermarking to be one of the lightweight solutions in limited networks such as \gls*{iot}. Hence, the features of data hiding that watermarking provides a new provenance encoding technique to store provenance records in an efficient way in-terms of storage and transmission overhead. Only one relevant work~\cite{Sultana2011} considered watermarking for data provenance. They consider inter-packet timing characteristics for embedding provenance information and does not use the data features and watermarking embedding with the data packets. This technique can be used to develop lightweight solutions for the limitations of IoT networks in terms of storage and computation.
    \item \textbf{\textit{Lack of complete provenance management system and narrow focus:}} The literature is still missing a robust provenance management system that take into account all required procedures from collecting, storing and analyzing provenance information. The system should satisfy the security requirements while maintaining storage and data management issues. To achieve this, these systems should be designed based on the architecture of \gls*{iot} networks, taking into account various factors, such as communication protocols, computational capabilities, storage limitations, node coordination, database integration, and potential security attacks against data and provenance records. Additionally, it is essential to thoroughly test the robustness of the proposed provenance system, assessing its performance across all relevant metrics. This approach ensures that the system complies with all the necessary objectives, rather than evaluating its performance based on only a subset of criteria. Many of the existing approaches fail to adequately address these important issues while maintaining the security of provenance data. Furthermore, the proposed provenance approaches focus on a specific application for a specific domain. The fast development of interconnected applications of different domains require adapting these techniques to be applied to multiple domains that include different structure of provenance records.
    \item \textbf{\textit{Lack of efficient query and provenance support:}} In \gls*{iot} networks, which involve numerous interconnected devices generating huge amounts of data, efficiently querying and managing the history and transformations of this data is a complex task. The absence of effective mechanisms for querying and handling provenance can hinder the ability to trace the origin and processing of data, resulting in a limitation in providing security, freshness, and scalability in \gls*{iot} applications. Addressing this issue involves developing more efficient and scalable methods for querying and managing provenance data in the unique and dynamic environment of \gls*{iot} networks.
    \item \textbf{\textit{The storage of blockchain is expensive:}} Blockchain relies on a decentralized and distributed network of nodes, each maintaining a copy of the entire blockchain. While this redundancy enhances security and fault tolerance, it significantly increases the storage requirements. In \gls*{iot} networks, with a large number of devices generating data, the continuous increase in the size of the blockchain can grow rapidly. Additionally, achieving scalability in blockchain networks, especially in public networks, is a known challenge. As the number of transactions and participants increases, maintaining the performance and efficiency of the blockchain becomes a complex task, hence requiring more robust infrastructure and resulting in an increased operational costs. In order to overcome the issue of expensive storage in blockchain-based data provenance for \gls*{iot} networks, researchers should develop a more scalable and efficient validation mechanisms, optimization of data storage techniques, and alternative blockchain architectures in \gls*{iot} that balance the advantages of security with the need for cost-effective scalability in the context of \gls*{iot} data provenance.
     \item \textbf{\textit{Integrating data provenance with \gls*{ids}:}} Adding data provenance to system logs transforms them from simple records into rich sources of security insights which are used in traditional Host-based Intrusion Detection Systems (HIDS). This helps investigators to detect security threats with greater accuracy, minimizing computational time and resources on false alarms, as extensive research has documented~\cite{ayoade2020,barre2019,berrada2019,xie2016,xie2020,han2017,du2017,gao2018,xie2021,berrada2020,peng2018,mathieu2020,hassan2020,QI2020}. Data provenance can be integrated with HIDS to improve intrusion detection. \gls*{ids} based on provenance, known as Provenance-based Intrusion Detection Systems (PIDS), leverage data provenance to identify intrusions. This involves examining not only the properties of system entities but also dealing with the causal relationships and information flow within a provenance graph~\cite{zipperele2022}. However, the integration of data provenance and IDS in IoT networks has not been widely studied. This opens an important field for researchers to study, addressing the significance of integrating data provenance in IoT networks and exploring its various applications for enhancing intrusion detection.
\end{itemize}

\section{Conclusion}
\label{sec:conclusion}
This review explores the integration of \gls*{iot} and data provenance, addressing the vulnerabilities in \gls*{iot} networks and the need to ensure data trustworthiness, data quality, traceability, and security. The \gls*{iot}, with its huge network of interconnected physical objects, has engaged applications in many domains, but its risk of cyber attacks is a growing concern. This paper aims to fill the gap by analyzing the integration of data provenance with \gls*{iot} networks which is considered as a security measure to securely transmit \gls*{iot} data and ensure data trustworthiness, a concept originally used in heterogeneous database systems. The core objective of our research was to provide a detailed review of existing data provenance techniques in \gls*{iot} networks, presenting their advantages, limitations, and their application in ensuring security requirements. We also examined the encountered attacks and performance metrics used to evaluate security requirements and system efficiency. Through a well structured taxonomy, we categorized different attributes related to the development of data provenance in \gls*{iot}, helping researchers for a better understanding of this field.

The findings of this study provide the importance of addressing the security challenges in an environment characterized by rapid data transmission, limited storage capabilities, bandwidth constraints, and energy limitations in \gls*{iot} networks. We have highlighted the need for secure, efficient and practical implementation of data provenance techniques. Moreover, we have identified that, while data provenance has been extensively studied in many domains, it remains an under explored area in the context of \gls*{iot} networks. The lack of a systematic literature review in this field motivated us to conduct this research, serving as a valuable resource for researchers. We have also addressed the following research questions: (1) How data provenance is linked to \gls*{iot} networks?; (2) What are the provenance storage techniques, attacks and security requirements when integrating provenance with \gls*{iot}?; (3) How can data provenance security approaches for \gls*{iot} networks be categorized?; (4) What are the existing practical implementations of data provenance in \gls*{iot}?; (5) What are the advantages and limitations of the proposed techniques for data provenance in \gls*{iot} in the studied literature? and (6) What are the main research gaps and challenges in the domain of data provenance in \gls*{iot} networks?. In our discussion, we have explored the differences between our review and the related research, presenting unresolved issues, research challenges and potential directions for future research. Hence, we have addressed both the significant challenges and the evolving issues within data provenance in \gls*{iot}. 

\bibliographystyle{ACM-Reference-Format}
\bibliography{main.bib}

\newpage

\begin{appendix}

\section{Additional Figures and Comparison Tables}
\label{appendix}

\begin{figure*}[hptb]
  \centering
  \includegraphics[width=.7\textwidth]{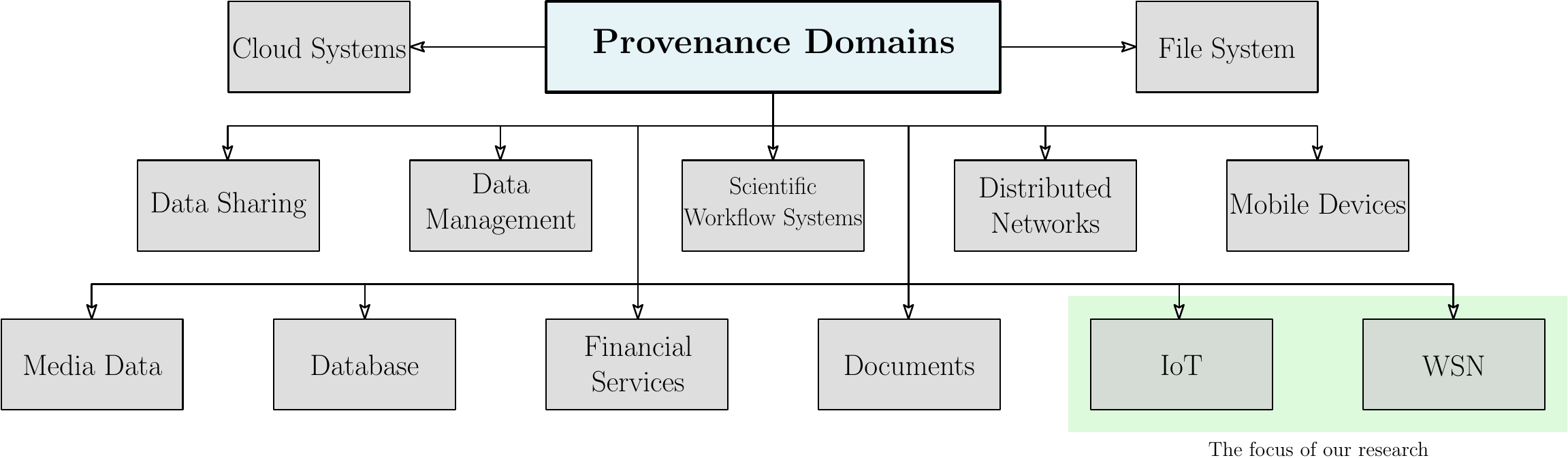}
  \caption{Provenance domains and the scope of our research. }
  \label{fig:taxonomyfocus}
\end{figure*}

\noindent Data provenance is presented in many domains such as file systems, databases, cloud computing, \gls*{iot}, distributed networks and many more domains. This research focuses on data provenance within the \gls*{iot} and \gls*{wsn} domains, as shown in Figure~\ref{fig:taxonomyfocus}.\\

\begin{figure}[!hptb]
  \centering
  \includegraphics[width=.7\textwidth]{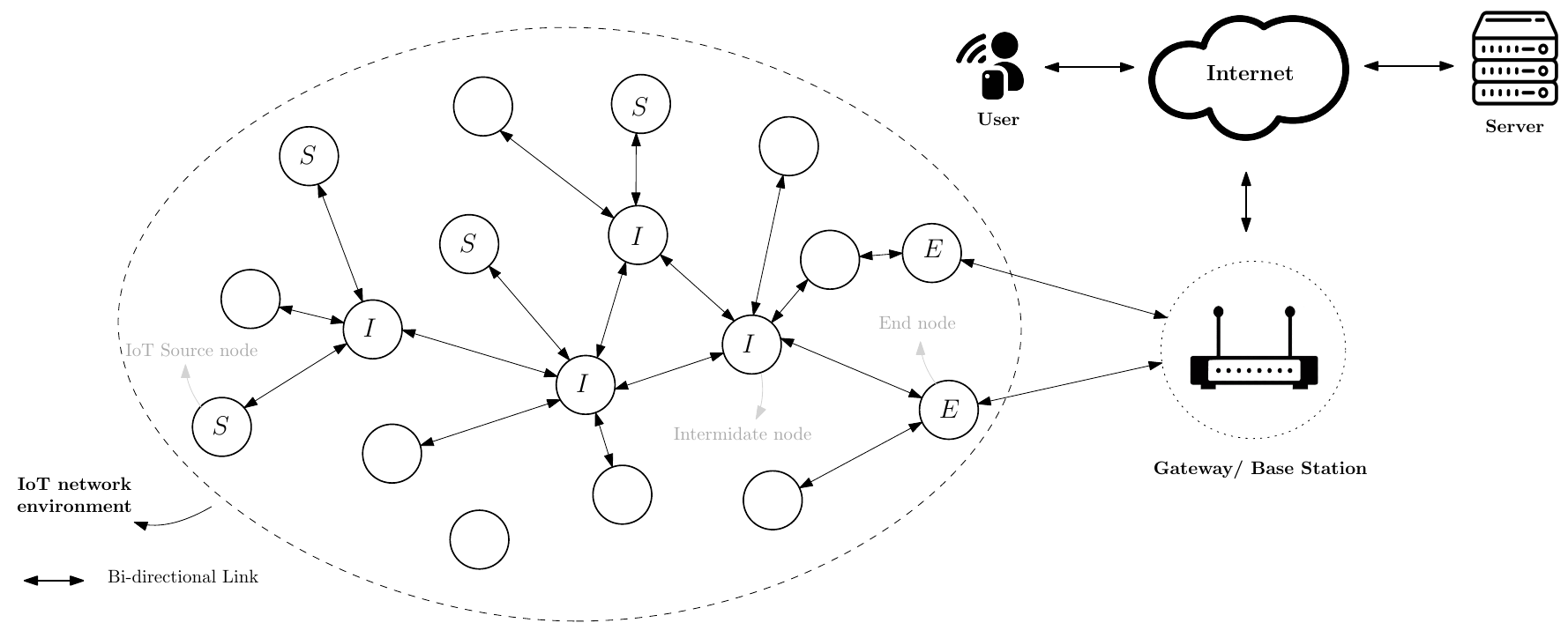}
  \caption{IoT network model assumed in our work.}
  \label{fig:network}
\end{figure}

\noindent Internet of Things (IoT) networks consist of a number of sensing nodes that communicate using a wireless multi-hop model, as shown in Figure~\ref{fig:network}. \\

\begin{figure*}[!hptb]
  \centering
  \includegraphics[width=0.8\textwidth]{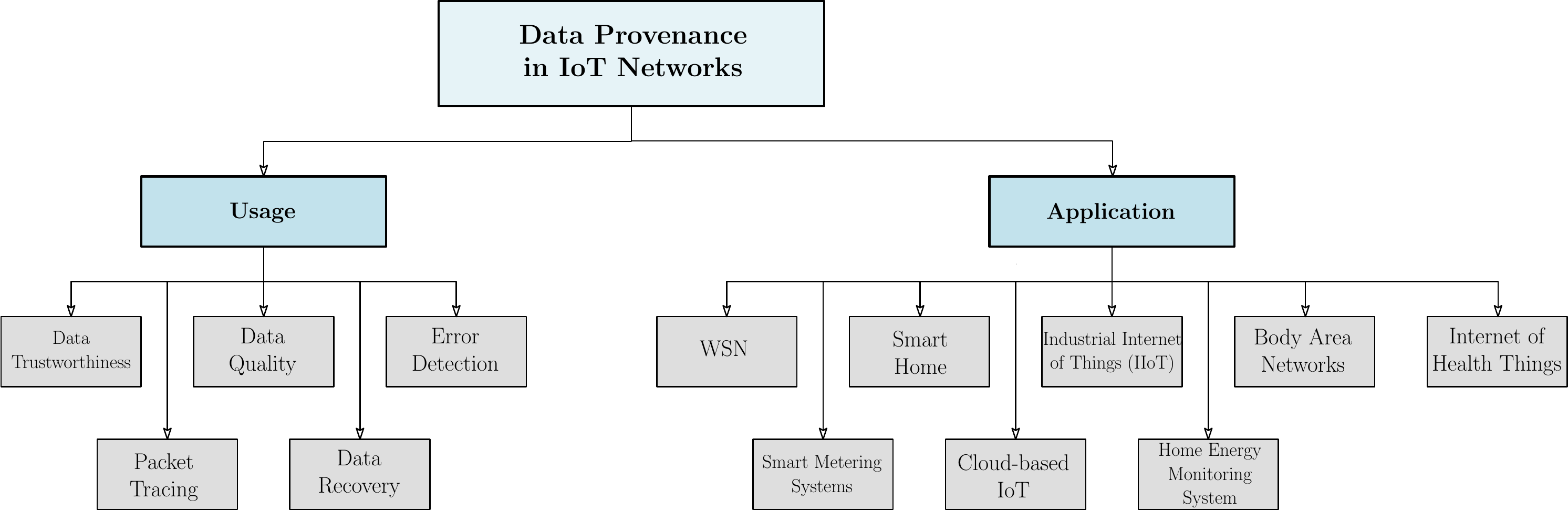}
  \caption{Taxonomy of the usage and applications for security techniques in IoT networks.}
  \label{fig:taxonomy_2}
\end{figure*}

Figure~\ref{fig:taxonomy_2} provides a taxonomy of the usage and applications for security techniques in IoT networks.

\begin{table*}[!htbp]
\begin{scriptsize}
\renewcommand{\arraystretch}{1.5}
\caption{Provenance category and evaluation methodology used in the selected security techniques.}
\begin{center}
\begin{tabular}{m{2.5cm} | m{1.5cm} | m{2.1cm} | m{1.2cm} | m{2.5cm} | m{4.2cm}}

\hline
\multirow{2}{*}{\textbf{Reference}} & \multirow{2}{*}{\textbf{Provenance}}  & \multicolumn{2}{c}{\textbf{Implementation}} & \multicolumn{2}{|c}{\textbf{Simulation}} \\ \cline{3-6}

& \textbf{Category} & \textbf{Hardware} & \textbf{Cloud-based} & \textbf{Programming Language} & \textbf{Simulator} \\ 
\hline
\hline
~\citet{Lim2010} &  Cryptography & \xmark & \xmark & Java Program & \xmark \\
\hline
~\citet{wang20162} & Cryptography & TelosB motes & \xmark & \xmark  & TinyOS/TOSSIM \\ 
\hline
~\citet{Chia2017} & Cryptography & Raspberry Pi 3/ DLink Smart Plug & \xmark & Java Program & \xmark \\
\hline
~\citet{suhail2018} & Cryptography & \xmark &  \xmark & \xmark & Contiki simulator, Cooja ~\cite{osterlind2006} \\
\hline
~\citet{xu2019} & Cryptography & Zigbee sensor nodes & \xmark & \xmark  & TinyOS/ PowerTOSSIMz\\
\hline
~\citet{SUHAIL2020} & Cryptography & \xmark  & \xmark   & \xmark & Contiki simulator, Cooja\\
\hline
~\citet{liu2020} & Cryptography & \xmark & \xmark & Python Program & \xmark \\
\hline
~\citet{tang2020} & Cryptography & \xmark  & \xmark & \xmark & \xmark \\
\hline
~\citet{Xu2022} & Cryptography & Zigbee/ TI CC2530 microcontroller & \xmark  & \xmark & TinyOS PowerTOSSIMz~\cite{perla2008} \\
\hline
~\citet{xu20222} & Cryptography & ZigBee/ TI CC2530 & \xmark & \xmark & TinyOS/ PowerTOSSIMz \\
\hline
~\citet{Salmin2015} & Bloom Filter & \xmark  & \xmark & \xmark & TinyOS simulator \\
\hline
~\citet{SIDDIQUI2019} & Bloom Filter & \xmark  & \xmark & Java Program & \xmark  \\
\hline
~\citet{harshan2020} & Bloom Filter & Raspberry Pi 3+ and a Digi XBee S2C & \xmark & \xmark & \xmark  \\
\hline
~\citet{syed2014} & Fingerprints & MICAz motes, running TinyOS & \xmark & \xmark & \xmark \\
\hline
~\citet{ALAM2014} & Fingerprints & TelosB motes & \xmark & \xmark & TOSSIM \\
\hline
~\citet{Kamal2018} & Fingerprints & MICAz motes  & \xmark  & \xmark & \xmark \\
\hline
~\citet{Nathalie2017} & Blockchain & \xmark & \xmark & \xmark & \xmark\\
\hline
~\citet{zeng2018} & Blockchain & MICAz motes & \xmark & \xmark & TOSSIM/TinyOS 2.1.2  \\
\hline
~\citet{Javaid2018} & Blockchain & \xmark & \xmark & \xmark & Solidity IDE, Remix  \\
\hline
~\citet{Sigwart2020} & Blockchain & \xmark & \xmark & \xmark & Ethereum (Rinkeby and Ropsten) \\
\hline
~\citet{liu20202} & Blockchain & \xmark  & \xmark & \xmark & Python interface/Pinocchio~\cite{Parno2016} and libsnark interface~\cite{kosba2018} with Etheruem test network \\
\hline
~\citet{Siddiqui2020} & Blockchain & \xmark & \xmark & \xmark & Hyperledger Fabric~\cite{christian2016}/ Java (BFT-SMaRt)~\cite{Sousa2018} \\
\hline
~\citet{Porkodi2021} & Blockchain & \xmark & \xmark & \xmark & N/A  \\
\hline
~\citet{yin2022} & Blockchain & \xmark & \xmark & \xmark & Ethereum network test/ Ubuntu \\
\hline
~\citet{sun2022} & Blockchain & \xmark & \xmark & \xmark & Public \gls*{iot} database MMASH~\cite{rossi2020} \\
\hline
~\citet{chenli2022} & Blockchain & \xmark & \xmark & Go language  and Hyperledger Fabric v2.0 & \xmark\\ 
\hline
~\citet{Sultana2011} & Watermarking & \xmark & \xmark & \xmark & TinyOS simulator  \\
\hline
~\citet{aman2019} & \gls*{pufs} & MICAz motes/ MATLAB & \xmark & \xmark & MICA 2 mote platform/ AVRORA  \\
\hline
~\citet{aman2021} & \gls*{pufs} & MICAz motes, CC2420 transceiver  & \xmark & \xmark & \xmark  \\ 
\hline
~\citet{Hamadeh2021} & \gls*{pufs} & FPGA Altera Cyclon & \xmark & \xmark & Verilog (HDL), ModelSim XE \\
\hline
~\citet{chang2022} & Graph-based & \xmark & \xmark & \xmark & Gentle Wake-Up/ SmartThings IDE~\cite{SmartThings2023,Samsung2023} \\
\hline
~\citet{Jaigirdar2023} & Graph-based & \xmark & \xmark & Flask module in python/ Neo4j & \xmark\\
\hline
~\citet{Al-Rakhami2022} & Graph-based & \xmark & \xmark & Eclipse Paho \tablefootnote{https://www.eclipse.org/paho/} and Python  & \xmark  \\
\hline
~\citet{Lomotey2019} & Data Sanitization & \xmark  & \xmark & \xmark  & \xmark \\
\hline
~\citet{Lomotey2018} & Lexical chain & Bluetooth devices & Amazon EC2 platform &  \xmark & \xmark \\
\hline
~\citet{gao2013} & Path difference & \xmark  & \xmark & \xmark & TinyOS with CTP~\cite{Gnawali2009}/ TOSSIM\\
\hline
~\citet{SADINENI2023} & Logging-based & \xmark & \xmark & \xmark & Contiki simulator, Cooja \\
\hline
~\citet{Alkhalil2019} & Fuzzy logic  & \xmark & \xmark & Java program & \xmark\\
\hline
~\citet{Wang2018} & Storing & \xmark & \xmark  & \xmark & Samsung SmartThings IDE~\cite{Smart2017}  \\
\hline
~\citet{elkhodr2020} & Storing & \xmark & \xmark & \xmark & \xmark \\
\hline

\end{tabular}
\label{tab:category}
\end{center}
\end{scriptsize}
\end{table*}

\begin{table*}
\begin{scriptsize}
\renewcommand{\arraystretch}{1.5}
\caption{Performance metrics used in the selected studies.}
\begin{center}
\begin{tabular}{c | c c c c c c c c}

\hline
\multirow{2}{*}{\textbf{Reference}} & \textbf{Prov.} & \textbf{Energy} & \textbf{Data Packet} & \textbf{Link-Loss} & \textbf{Detection} & \textbf{False} & \textbf{False}  & \textbf{Computation} \\ 

 & \textbf{Length} & \textbf{Consumption} & \textbf{Size} & \textbf{Rate} & \textbf{Rate} & \textbf{Positive Rate} & \textbf{Negative Rate} & \textbf{time} \\ 
\hline
\hline
~\citet{Lim2010} & \xmark & \xmark & \xmark & \xmark & \xmark & \xmark & \xmark & \usym{1F5F8} \\
~\citet{wang20162} & \usym{1F5F8} & \usym{1F5F8} & \xmark & \usym{1F5F8} & \usym{1F5F8} & \xmark & \xmark &  \usym{1F5F8} \\
~\citet{Chia2017} & \xmark & \xmark & \xmark & \xmark & \xmark & \xmark & \xmark & \usym{1F5F8} \\
~\citet{suhail2018} & \xmark & \usym{1F5F8} & \xmark & \xmark & \xmark & \xmark & \xmark & \xmark \\
~\citet{xu2019} & \usym{1F5F8} & \usym{1F5F8} & \usym{1F5F8} & \xmark & \xmark & \xmark & \xmark & \xmark \\
~\citet{SUHAIL2020} & \usym{1F5F8} & \usym{1F5F8} & \xmark & \xmark & \usym{1F5F8} & \xmark & \xmark & \usym{1F5F8} \\
~\citet{liu2020} & \usym{1F5F8} & \xmark & \xmark & \xmark & \usym{1F5F8} & \usym{1F5F8} & \usym{1F5F8} & \xmark \\
~\citet{tang2020} & \xmark & \xmark & \xmark & \xmark & \xmark & \xmark & \xmark & \xmark\\
~\citet{Xu2022} & \usym{1F5F8} & \usym{1F5F8} & \usym{1F5F8} & \xmark & \xmark & \usym{1F5F8} & \xmark & \xmark \\
XXu and Wang~\cite{xu20222} & \usym{1F5F8} & \usym{1F5F8} & \usym{1F5F8} & \xmark & \usym{1F5F8} & \xmark  & \xmark & \usym{1F5F8}  \\
~\citet{Salmin2015} & \usym{1F5F8} & \usym{1F5F8} & \xmark & \usym{1F5F8} & \usym{1F5F8} & \usym{1F5F8} & \xmark & \xmark \\
~\citet{SIDDIQUI2019} & \xmark & \xmark & \xmark & \xmark & \xmark & \xmark & \xmark & \usym{1F5F8} \\
~\citet{harshan2020} & \usym{1F5F8} & \xmark & \xmark & \xmark & \usym{1F5F8} & \xmark & \xmark & \usym{1F5F8} \\
~\citet{syed2014} & \xmark & \usym{1F5F8} & \usym{1F5F8} & \xmark & \xmark & \xmark & \xmark & \usym{1F5F8} \\
~\citet{ALAM2014} & \usym{1F5F8} & \usym{1F5F8} & \usym{1F5F8} & \xmark & \xmark & \xmark & \xmark & \xmark \\
~\citet{Kamal2018} & \xmark & \usym{1F5F8} & \xmark & \xmark & \xmark & \xmark & \xmark & \usym{1F5F8} \\
~\citet{Nathalie2017} & \xmark & \xmark & \xmark & \xmark & \xmark & \xmark & \xmark & \xmark \\
~\citet{zeng2018} & \usym{1F5F8} & \xmark & \usym{1F5F8} & \xmark & \xmark & \xmark & \xmark & \xmark\\
~\citet{Javaid2018} & \xmark & \xmark & \xmark & \xmark & \xmark & \xmark  & \xmark & \xmark  \\
~\citet{Sigwart2020} & \xmark & \xmark & \xmark & \xmark & \xmark & \xmark & \xmark & \usym{1F5F8} \\
~\citet{liu20202} & \usym{1F5F8} & \xmark & \usym{1F5F8} & \xmark & \xmark & \xmark & \xmark & \usym{1F5F8} \\
~\citet{Siddiqui2020} & \xmark & \xmark & \xmark & \xmark & \xmark & \xmark & \xmark & \usym{1F5F8} \\
~\citet{Porkodi2021} & \xmark & \xmark & \xmark & \xmark & \xmark & \xmark & \xmark & \usym{1F5F8} \\
~\citet{yin2022} & \xmark & \xmark & \xmark & \xmark & \xmark & \xmark & \xmark & \usym{1F5F8} \\
~\citet{sun2022} & \xmark & \xmark & \xmark & \xmark & \xmark & \xmark & \xmark & \usym{1F5F8} \\
~\citet{chenli2022} & \xmark & \xmark & \xmark & \xmark & \xmark & \xmark & \xmark &  \usym{1F5F8} \\
~\citet{Sultana2011} & \xmark & \xmark & \usym{1F5F8} & \xmark & \usym{1F5F8} & \usym{1F5F8} & \usym{1F5F8} & \xmark \\
~\citet{aman2019} & \usym{1F5F8} & \usym{1F5F8} & \usym{1F5F8} & \xmark & \xmark & \xmark & \xmark & \usym{1F5F8}\\
~\citet{aman2021} & \usym{1F5F8} & \usym{1F5F8} & \xmark & \xmark & \xmark & \xmark & \xmark & \xmark \\
~\citet{Hamadeh2021} & \xmark & \xmark & \xmark & \xmark & \xmark & \xmark & \xmark & \usym{1F5F8} \\
~\citet{chang2022} & \xmark & \xmark & \xmark & \xmark & \xmark & \xmark  & \xmark & \usym{1F5F8} \\
~\citet{Jaigirdar2023} & \usym{1F5F8} & \xmark & \xmark & \xmark & \xmark & \xmark & \xmark & \usym{1F5F8} \\
~\citet{Al-Rakhami2022} & \xmark & \xmark & \usym{1F5F8} & \xmark & \xmark & \xmark & \xmark & \usym{1F5F8} \\
~\citet{Lomotey2019} & \xmark & \xmark & \xmark & \xmark & \xmark & \xmark & \xmark & \xmark \\
~\citet{Lomotey2018} & \xmark & \xmark & \xmark & \xmark & \usym{1F5F8} & \usym{1F5F8} & \usym{1F5F8} & \usym{1F5F8} \\
~\citet{gao2013} & \usym{1F5F8} & \xmark & \usym{1F5F8} & \usym{1F5F8} & \xmark & \usym{1F5F8} & \xmark & \usym{1F5F8} \\
~\citet{SADINENI2023} & \usym{1F5F8} & \xmark & \usym{1F5F8} & \xmark & \xmark & \xmark & \xmark & \xmark \\
~\citet{Alkhalil2019} & \xmark & \xmark & \xmark & \xmark & \xmark & \xmark & \xmark & \usym{1F5F8} \\
~\citet{Wang2018} & \usym{1F5F8} & \xmark & \usym{1F5F8} & \usym{1F5F8} & \xmark & \xmark & \xmark & \usym{1F5F8} \\
~\citet{elkhodr2020} & \xmark & \xmark & \xmark & \xmark & \xmark & \xmark & \xmark & \xmark\\

\hline
\end{tabular}
\label{tab:performance}
\end{center}
\end{scriptsize}
\end{table*}

\begin{table*}[!htbp]
\begin{scriptsize}
\renewcommand{\arraystretch}{1.5}
\caption{Security requirements in the selected studies.}
\begin{center}
\begin{tabular}{c | c c c c c c c}

\hline
\multirow{2}{*}{\textbf{Reference}} & \multicolumn{7}{c}{\textbf{Security Requirements}} \\ 
\cline{2-8}
 & \textit{\textbf{Data Integrity}} & \textit{\textbf{Confidentiality}} & \textit{\textbf{Availability}} & \textit{\textbf{Privacy}} & \textit{\textbf{Freshness}} & \textit{\textbf{Non-repudiation}} & \textit{\textbf{Unforgeability}} \\ 
\hline
\hline
~\citet{Lim2010} & \xmark & \xmark & \usym{1F5F8}& \xmark & \usym{1F5F8}& \xmark & \xmark \\
~\citet{wang20162} & \usym{1F5F8}& \usym{1F5F8}& \usym{1F5F8}& \xmark & \usym{1F5F8}& \usym{1F5F8}& \xmark \\
~\citet{Chia2017} & \usym{1F5F8}& \xmark & \xmark & \xmark & \xmark & \usym{1F5F8}& \usym{1F5F8}\\
~\citet{suhail2018} & \usym{1F5F8}& \xmark & \xmark & \xmark & \usym{1F5F8}& \usym{1F5F8}& \usym{1F5F8}\\
~\citet{xu2019} & \usym{1F5F8}& \xmark & \xmark & \xmark & \xmark & \usym{1F5F8}& \xmark \\
~\citet{SUHAIL2020} & \xmark & \xmark & \usym{1F5F8}& \xmark & \usym{1F5F8}& \usym{1F5F8}& \xmark \\
~\citet{liu2020} & \usym{1F5F8}& \xmark & \usym{1F5F8}& \xmark & \usym{1F5F8} & \xmark & \usym{1F5F8}  \\
~\citet{tang2020} & \usym{1F5F8} & \usym{1F5F8} & \xmark & \xmark & \xmark & \xmark & \usym{1F5F8} \\
~\citet{Xu2022} & \usym{1F5F8} & \usym{1F5F8} & \usym{1F5F8} & \xmark & \xmark & \usym{1F5F8} & \usym{1F5F8}\\
~\citet{xu20222} & \usym{1F5F8} & \xmark & \usym{1F5F8} & \xmark & \xmark & \usym{1F5F8} & \usym{1F5F8}  \\
~\citet{Salmin2015} & \usym{1F5F8} & \usym{1F5F8} & \usym{1F5F8} & \xmark & \usym{1F5F8} & \xmark & \xmark  \\
~\citet{SIDDIQUI2019} & \usym{1F5F8} & \usym{1F5F8} & \usym{1F5F8} & \xmark & \xmark & \usym{1F5F8} & \xmark  \\
~\citet{harshan2020} & \usym{1F5F8} & \xmark & \usym{1F5F8} & \xmark & \xmark & \usym{1F5F8} & \usym{1F5F8} \\
~\citet{syed2014} & \usym{1F5F8} & \usym{1F5F8} & \xmark & \xmark & \usym{1F5F8} & \usym{1F5F8} & \usym{1F5F8}  \\
~\citet{ALAM2014} & \xmark & \xmark & \xmark & \xmark & \xmark & \xmark & \xmark \\
~\citet{Kamal2018} & \usym{1F5F8} & \xmark & \usym{1F5F8} & \xmark & \usym{1F5F8} & \xmark & \xmark  \\
~\citet{Nathalie2017} & \usym{1F5F8} & \usym{1F5F8} & \xmark & \usym{1F5F8} & \xmark & \usym{1F5F8} & \xmark \\
~\citet{zeng2018} & \usym{1F5F8} & \usym{1F5F8} & \xmark & \xmark & \usym{1F5F8} & \usym{1F5F8} & \usym{1F5F8} \\
~\citet{Javaid2018} & \usym{1F5F8} & \xmark & \xmark & \xmark & \usym{1F5F8} & \usym{1F5F8} & \xmark \\
~\citet{Sigwart2020} & \usym{1F5F8} & \xmark & \usym{1F5F8} & \xmark & \xmark & \usym{1F5F8} & \xmark \\
~\citet{liu20202} & \usym{1F5F8} & \xmark & \xmark & \xmark & \usym{1F5F8} & \usym{1F5F8} & \usym{1F5F8} \\
~\citet{Siddiqui2020} & \usym{1F5F8} & \usym{1F5F8} & \xmark & \xmark & \usym{1F5F8} & \usym{1F5F8} & \xmark \\
~\citet{Porkodi2021} & \usym{1F5F8} & \usym{1F5F8} & \usym{1F5F8} & \xmark & \xmark & \xmark & \usym{1F5F8}  \\
~\citet{yin2022} & \usym{1F5F8} & \usym{1F5F8} & \xmark & \usym{1F5F8} & \xmark & \usym{1F5F8} & \xmark \\
~\citet{sun2022} & \usym{1F5F8} & \usym{1F5F8} & \xmark & \xmark & \xmark & \xmark & \usym{1F5F8} \\
~\citet{chenli2022} & \usym{1F5F8} & \usym{1F5F8} & \xmark & \xmark & \xmark & \xmark & \xmark \\
~\citet{Sultana2011} & \usym{1F5F8} & \usym{1F5F8} & \usym{1F5F8} & \xmark & \usym{1F5F8} & \xmark & \xmark  \\
~\citet{aman2019} & \usym{1F5F8} & \usym{1F5F8} & \xmark & \usym{1F5F8} & \usym{1F5F8} & \usym{1F5F8} & \xmark\\
~\citet{aman2021} & \usym{1F5F8} & \xmark & \xmark & \xmark & \usym{1F5F8} & \xmark & \xmark  \\
~\citet{Hamadeh2021} & \usym{1F5F8} & \xmark & \usym{1F5F8} & \usym{1F5F8} & \usym{1F5F8} & \usym{1F5F8} & \xmark  \\
~\citet{chang2022} & \xmark & \xmark & \xmark & \xmark & \usym{1F5F8} & \xmark & \xmark \\
~\citet{Jaigirdar2023} & \usym{1F5F8} & \xmark & \usym{1F5F8} & \xmark & \xmark & \usym{1F5F8} & \usym{1F5F8} \\
~\citet{Al-Rakhami2022} & \usym{1F5F8} & \usym{1F5F8} & \usym{1F5F8} & \xmark & \xmark & \xmark & \usym{1F5F8} \\
~\citet{Lomotey2019} & \xmark & \usym{1F5F8} & \xmark & \usym{1F5F8} & \xmark & \usym{1F5F8} & \xmark \\
~\citet{Lomotey2018} & \usym{1F5F8} & \usym{1F5F8} & \usym{1F5F8} & \xmark & \xmark & \usym{1F5F8} & \xmark \\
~\citet{gao2013} & \xmark & \xmark & \xmark & \xmark & \usym{1F5F8} & \usym{1F5F8} & \xmark \\
~\citet{SADINENI2023} & \xmark & \xmark & \usym{1F5F8} & \xmark & \xmark & \xmark & \usym{1F5F8} \\
~\citet{Alkhalil2019} & \xmark & \xmark & \usym{1F5F8} & \xmark & \xmark & \usym{1F5F8} & \xmark \\
~\citet{Wang2018} & \usym{1F5F8} & \xmark & \usym{1F5F8} & \usym{1F5F8} & \xmark & \usym{1F5F8} & \xmark \\
~\citet{elkhodr2020} & \xmark & \xmark & \xmark & \xmark & \usym{1F5F8} & \xmark & \xmark\\

\hline
\end{tabular}
\label{tab:requirnemnts}
\end{center}
\end{scriptsize}
\end{table*}

\begin{table*}[!htbp]
\begin{scriptsize}
\renewcommand{\arraystretch}{1.5}
\caption{Attacks studied in the selected security techniques.}
\begin{center}
\resizebox{\textwidth}{!}{\begin{tabular}{>{\centering}m{1.9cm} | c c c c c | c c c c >{\centering}m{1.1cm} c}

\hline
\multirow{4}{*}{\textbf{Reference}} & \multicolumn{5}{c|}{\textbf{Data Attacks}} & \multicolumn{6}{c}{\textbf{Provenance Attacks}} \\ 

\cline{2-12}
 & \textit{\textbf{Drop}} & \textit{\textbf{Replay}} & \textit{\textbf{Forgery}} & \textit{\textbf{Modification}} & \textit{\textbf{Eavesdrop}} & \textit{\textbf{Record drop}} & \textit{\textbf{Replay}} & \textit{\textbf{Forging}} & \textit{\textbf{Modification}} & \textit{\textbf{Chain tampering}} & \textit{\textbf{Inference}} \\ 
\hline
\hline
~\citet{Lim2010}  & \xmark & \xmark & \xmark & \xmark & \xmark & \xmark & \xmark & \xmark & \xmark & \xmark & \xmark  \\
~\citet{wang20162} & \usym{1F5F8} & \usym{1F5F8} & \usym{1F5F8} & \usym{1F5F8} & \xmark & \usym{1F5F8} & \usym{1F5F8} & \usym{1F5F8} & \usym{1F5F8} & \xmark &  \xmark \\
~\citet{Chia2017} & \usym{1F5F8} & \usym{1F5F8} & \usym{1F5F8} & \usym{1F5F8} & \xmark & \usym{1F5F8} & \usym{1F5F8} & \usym{1F5F8} & \usym{1F5F8} & \xmark & \xmark \\
~\citet{suhail2018} & \usym{1F5F8} & \usym{1F5F8} & \usym{1F5F8} & \xmark & \usym{1F5F8} & \usym{1F5F8} & \usym{1F5F8} & \usym{1F5F8} & \xmark & \xmark & \xmark \\
~\citet{xu2019} & \xmark & \xmark & \xmark & \xmark & \xmark & \xmark & \xmark & \xmark & \xmark & \xmark & \xmark   \\
~\citet{SUHAIL2020} & \usym{1F5F8} & \usym{1F5F8} & \xmark & \xmark & \xmark & \usym{1F5F8} & \usym{1F5F8} & \xmark & \usym{1F5F8} & \usym{1F5F8} & \xmark  \\
~\citet{liu2020} & \usym{1F5F8} & \usym{1F5F8} & \usym{1F5F8} & \usym{1F5F8} & \xmark & \xmark & \xmark  & \xmark & \xmark & \xmark & \xmark  \\
~\citet{tang2020} & \xmark & \xmark & \usym{1F5F8} & \usym{1F5F8} & \usym{1F5F8} & \xmark & \xmark & \xmark & \xmark & \xmark & \xmark \\
~\citet{Xu2022} & \usym{1F5F8} & \xmark & \usym{1F5F8} & \usym{1F5F8} & \xmark & \usym{1F5F8} & \xmark & \usym{1F5F8} & \usym{1F5F8} & \xmark & \usym{1F5F8}\\
~\citet{xu20222} & \usym{1F5F8} & \xmark & \xmark & \usym{1F5F8} & \xmark & \usym{1F5F8}  & \xmark & \xmark & \usym{1F5F8}  & \xmark & \xmark  \\
~\citet{Salmin2015} & \usym{1F5F8} & \usym{1F5F8} & \xmark & \xmark & \usym{1F5F8} & \usym{1F5F8} & \xmark & \xmark & \xmark & \xmark & \usym{1F5F8}   \\
~\citet{SIDDIQUI2019} & \usym{1F5F8} & \usym{1F5F8} & \xmark & \usym{1F5F8} & \xmark & \xmark & \xmark & \xmark & \xmark & \xmark & \xmark   \\
~\citet{harshan2020} & \xmark & \xmark & \usym{1F5F8} & \usym{1F5F8} & \usym{1F5F8} & \xmark & \xmark & \usym{1F5F8} & \usym{1F5F8} & \xmark & \usym{1F5F8}\\
~\citet{syed2014} & \xmark & \usym{1F5F8} & \usym{1F5F8} & \usym{1F5F8} & \usym{1F5F8} & \xmark & \usym{1F5F8}  & \usym{1F5F8} & \xmark & \xmark & \xmark \\
~\citet{ALAM2014} & \xmark & \usym{1F5F8} & \usym{1F5F8}  & \usym{1F5F8}  & \xmark & \xmark & \usym{1F5F8}  & \usym{1F5F8}  & \usym{1F5F8}  & \xmark & \xmark \\
~\citet{Kamal2018} & \xmark & \xmark & \usym{1F5F8} & \xmark & \xmark & \xmark & \xmark & \xmark & \xmark & \xmark & \xmark   \\
~\citet{Nathalie2017} & \xmark & \xmark & \xmark & \usym{1F5F8} & \xmark & \xmark & \xmark & \xmark & \usym{1F5F8} & \xmark & \xmark \\
~\citet{zeng2018} & \usym{1F5F8} & \xmark & \xmark & \xmark & \usym{1F5F8} & \xmark & \xmark & \xmark & \xmark & \xmark & \xmark  \\
~\citet{Javaid2018} & \usym{1F5F8} & \xmark & \usym{1F5F8} & \usym{1F5F8} & \xmark & \usym{1F5F8}  & \xmark & \usym{1F5F8} & \usym{1F5F8} & \xmark & \xmark   \\
~\citet{Sigwart2020} & \xmark & \xmark & \usym{1F5F8} & \usym{1F5F8} & \usym{1F5F8} & \xmark & \xmark & \usym{1F5F8} & \usym{1F5F8} & \xmark & \xmark\\
~\citet{liu20202} & \xmark & \xmark & \usym{1F5F8} & \usym{1F5F8} & \xmark & \xmark & \xmark & \usym{1F5F8} & \usym{1F5F8} & \xmark & \xmark \\
~\citet{Siddiqui2020} & \xmark & \xmark & \usym{1F5F8} & \usym{1F5F8} & \xmark & \xmark & \xmark & \usym{1F5F8} & \usym{1F5F8} & \xmark & \xmark \\
~\citet{Porkodi2021} & \xmark & \xmark & \usym{1F5F8} & \xmark & \xmark & \xmark & \xmark & \usym{1F5F8} & \xmark & \xmark & \xmark \\
~\citet{yin2022} & \xmark & \xmark & \usym{1F5F8} & \usym{1F5F8} & \xmark & \xmark & \xmark & \usym{1F5F8} & \usym{1F5F8}  & \xmark & \xmark  \\
~\citet{sun2022} & \xmark & \xmark & \usym{1F5F8} & \usym{1F5F8} & \xmark & \xmark & \xmark & \usym{1F5F8} & \usym{1F5F8} & \xmark & \xmark \\
~\citet{chenli2022} & \xmark & \xmark & \usym{1F5F8} & \usym{1F5F8} & \xmark & \xmark & \xmark & \usym{1F5F8} & \usym{1F5F8} & \xmark &  \xmark \\
~\citet{Sultana2011} & \usym{1F5F8} & \xmark & \xmark & \usym{1F5F8} & \xmark & \xmark & \xmark  & \xmark & \xmark & \xmark & \xmark  \\
~\citet{aman2019} & \usym{1F5F8} & \usym{1F5F8} & \xmark & \usym{1F5F8} & \usym{1F5F8} & \usym{1F5F8} & \usym{1F5F8} & \xmark & \usym{1F5F8} & \xmark & \usym{1F5F8}\\
~\citet{aman2021} & \usym{1F5F8} & \usym{1F5F8} & \usym{1F5F8} & \usym{1F5F8} & \xmark & \xmark & \xmark  & \xmark & \xmark & \xmark & \xmark  \\
~\citet{Hamadeh2021} & \xmark & \xmark & \usym{1F5F8} & \usym{1F5F8} & \xmark & \xmark & \xmark & \usym{1F5F8} & \usym{1F5F8} & \xmark & \xmark \\
~\citet{chang2022} & \xmark & \xmark & \xmark & \xmark & \xmark & \xmark  & \xmark & \xmark & \xmark & \xmark & \xmark  \\
~\citet{Jaigirdar2023} & \xmark & \xmark & \usym{1F5F8} & \usym{1F5F8} & \xmark & \xmark & \xmark & \usym{1F5F8} & \usym{1F5F8}  & \xmark & \xmark  \\
~\citet{Al-Rakhami2022} & \xmark & \xmark & \usym{1F5F8} & \usym{1F5F8} & \xmark & \xmark & \xmark & \usym{1F5F8} & \usym{1F5F8} & \xmark & \xmark \\
~\citet{Lomotey2019} & \xmark & \xmark & \xmark & \xmark & \xmark & \xmark & \xmark & \xmark & \xmark & \xmark & \xmark \\
~\citet{Lomotey2018} & \xmark & \xmark & \xmark & \xmark & \xmark & \xmark & \xmark & \xmark & \xmark & \xmark & \xmark \\
~\citet{gao2013} & \xmark & \xmark & \xmark & \xmark & \xmark & \xmark & \xmark & \xmark & \xmark  & \xmark & \xmark  \\
~\citet{SADINENI2023} & \usym{1F5F8} & \xmark & \usym{1F5F8} & \usym{1F5F8} & \xmark & \usym{1F5F8} & \xmark & \usym{1F5F8} & \usym{1F5F8}  & \xmark & \xmark  \\
~\citet{Alkhalil2019} & \xmark & \xmark & \xmark & \xmark & \xmark & \xmark & \xmark & \xmark & \xmark & \xmark & \xmark \\
~\citet{Wang2018} & \xmark & \xmark & \usym{1F5F8} & \usym{1F5F8} & \usym{1F5F8} & \xmark & \xmark & \usym{1F5F8} & \usym{1F5F8} & \usym{1F5F8} & \usym{1F5F8}\\
~\citet{elkhodr2020} & \xmark & \xmark & \xmark & \xmark & \xmark & \xmark & \xmark & \xmark & \xmark & \xmark & \xmark\\

\hline
\end{tabular}}
\label{tab:attacks}
\end{center}
\end{scriptsize}
\end{table*}
\end{appendix}

\end{document}